\documentclass[a4,12pt]{article}

\usepackage{bm}
\usepackage{graphicx}
\usepackage{amsmath}
\usepackage{amsfonts}

\makeatletter
  
  \@addtoreset{equation}{section}
\makeatother

\newcommand{\ol}{\overline}
\newcommand{\wt}{\widetilde}

\def\tr{\mathop{\rm tr}\nolimits}
\def\Pexp{\mathop{\rm Pexp}\nolimits}
\def\Hom{\mathop{\rm Hom}\nolimits}

\newcommand{\ZZ}{\mathbb{Z}}
\newcommand{\RR}{\mathbb{R}}

\newcommand{\nn}{\nonumber}

\begin{document}

\begin{titlepage}
\title{
\vspace{-1.5cm}
\begin{flushright}
{\normalsize TIT/HEP-675\\ Nov 2019}
\end{flushright}
\vspace{1.5cm}
\LARGE{Finite $N$ corrections to the superconformal index
of toric quiver gauge theories}}
\author{
Reona {\scshape Arai\footnote{E-mail: r.arai@th.phys.titech.ac.jp}},
Shota {\scshape Fujiwara\footnote{E-mail: s.fujiwara@th.phys.titech.ac.jp}},\\
Yosuke {\scshape Imamura\footnote{E-mail: imamura@phys.titech.ac.jp}},
and
Tatsuya {\scshape Mori\footnote{E-mail: t.mori@th.phys.titech.ac.jp}}
\\
\\
{\itshape Department of Physics, Tokyo Institute of Technology}, \\ {\itshape Tokyo 152-8551, Japan}}

\date{}
\maketitle
\thispagestyle{empty}

\begin{abstract}
The superconformal index of quiver gauge theories
realized on D3-branes in toric Calabi-Yau cones is investigated.
We use the AdS/CFT correspondence and study D3-branes wrapped on supersymmetric cycles.
We focus on brane configurations in which a single D3-brane is wrapped on a cycle,
and we do not take account of branes with multiple wrapping.
We propose a formula that gives finite $N$ corrections to the index
caused by such brane configurations.
We compare the predictions of the formula
for several examples
with the results on the gauge theory side
obtained by using localization
for small size of gauge groups,
and confirm
that the formula correctly reproduces the finite $N$ corrections
up to expected order.
\end{abstract}
\end{titlepage}
\tableofcontents

\section{Introduction}
The AdS/CFT correspondence \cite{Maldacena:1997re,Witten:1998qj,Gubser:1998bc} has been extensively investigated, and a lot of evidences have been found.
However, majority of the previous works is about
the large $N$ limit.
A reason for this is that as the parameter relation
\begin{align}
N\sim\frac{L^4}{l_p^4}
\label{param}
\end{align}
shows if $N$ is small the Planck length $l_p$ becomes
comparable with the AdS radius $L$ and the
quantum gravity effect is expected to be relevant.
Even so,
we may be able to obtain non-trivial evidences for
the AdS/CFT correspondence with finite $N$ by analyzing
quantities protected by supersymmetry.
In this paper we use the superconformal index \cite{Kinney:2005ej}
as such a protected quantity, and study the AdS/CFT correspondence
for finite $N$.
Even if the quantum gravity correction does not affect the index
there is another source of finite $N$ corrections.
The parameter relation (\ref{param})
can be rewritten in terms of the D3-brane tension $T_{\rm D3}\sim l_p^{-4}$ as
\begin{align}
N\sim L^4T_{\rm D3},
\end{align}
and we can regard $N$ as a typical value of the classical action
of a D3-brane wrapped on a cycle with scale $L$.
Namely, D3-branes wrapped on supersymmetric cycles
in the internal space
contribute to the superconformal index as a finite $N$ correction.
Such wrapped branes were first studied in \cite{Witten:1998xy} in the context of
the AdS/CFT correspondence,
and found to correspond to
baryonic operators on the gauge theory side.
The purpose of this paper is to
investigate this correspondence
at the level of the superconformal index.
We have already performed similar analyses
for S-fold theories \cite{Arai:2019xmp} and orbifold quiver gauge theories \cite{Arai:2019wgv}.
In this paper we consider toric quiver gauge theories as
a natural extension of them.

A toric quiver gauge theory is defined as a gauge theory
realized by putting D3-branes in a toric Calabi-Yau three fold.
There is a systematic prescription \cite{Feng:2000mi,Feng:2002zw}
to determine the gauge theory from the toric data of the Calabi-Yau.
The holographic dual is type IIB string theory in $AdS_5\times SE_5$,
where $SE_5$ is the Sasaki-Einstein manifold which is the base of the Calabi-Yau cone.
This relation provides infinite number of examples of the AdS/CFT correspondence.

On the gauge theory side, in principle,
the index can be calculated with the help of localization formula,
which includes the matrix integral.
For a small class of theories including
the ${\cal N}=4$ supersymmetric $U(N)$ Yang-Mills theory
and ${\cal N}=2$ circular quiver gauge theories
the matrix integral in the Schur limit was carried out and
analytic formulas were obtained \cite{Bourdier:2015wda,Bourdier:2015sga}.
However, in general, we need to rely on the numerical analysis,
and it is practically
difficult to obtain the index up to desired order
for a gauge group with large rank.
In this paper we consider only theories with $N=2$ and $N=3$,
for which we can calculate the index within realistic time.

On the gravity side we calculate the finite $N$ correction
as the contribution of wrapped D3-branes.
In this paper we take account of brane configurations
with single wrapping.
Although configurations with multiple wrapping would also contribute to the index,
they give higher order terms than single-wrapping ones, and we neglect them in this paper.
Concretely, we propose the following relation
\begin{align}
{\cal I}={\cal I}^{\rm KK}\left(1+\sum_I{\cal I}_{S_I}^{\rm D3}+\cdots\right),
\label{adsproposal}
\end{align}
where ${\cal I}_{S_I}^{\rm D3}$
is the contribution of a D3-brane wrapped on
a supersymmetric three-cycle $S_I$
corresponding to a corner of the toric diagram.
The dots represent the contribution of multiple wrapping
which is out of scope of this paper.

This paper is organized as follows.
In the next section we summarize basic aspects of toric
geometry and the corresponding quiver gauge theories.
We define $U(1)$ charges for a general
toric quiver gauge theories
and define the superconformal index in terms of them.
In Section \ref{ads.sec} we propose an explicit expression for the
D3-brane contribution ${\cal I}_{S_I}^{\rm D3}$ in (\ref{adsproposal}).
We give a simple prescription to calculate it
for each supersymmetric cycle from the toric data depicted as the
toric diagram.
In Section \ref{examples.sec} we confirm the correctness of the formula
by calculating the index on the both sides of the AdS/CFT correspondence.
We show the results for $N=2$ in the main text, and results for
$N=3$ in Appendix.
In Section \ref{discussion.sec} we summarize our results.
In Appendix we give additional results and explanations.

\section{Toric geometry and quiver gauge theories}\label{toric.sec}
\subsection{Toric data}
By definition a toric Calabi-Yau $n$-fold has $U(1)^n$ isometry,
and we can describe it as $\bm{T}^n$ fibration over $n$-dimensional base $B$,
which is a closed subset of $\wt{\cal V}=\RR^n$.
The action of $U(1)^n$ is not free
and the fiber reduces to lower-dimensional torus on the boundary of $B$.

\begin{figure}[htb]
\centering
\includegraphics{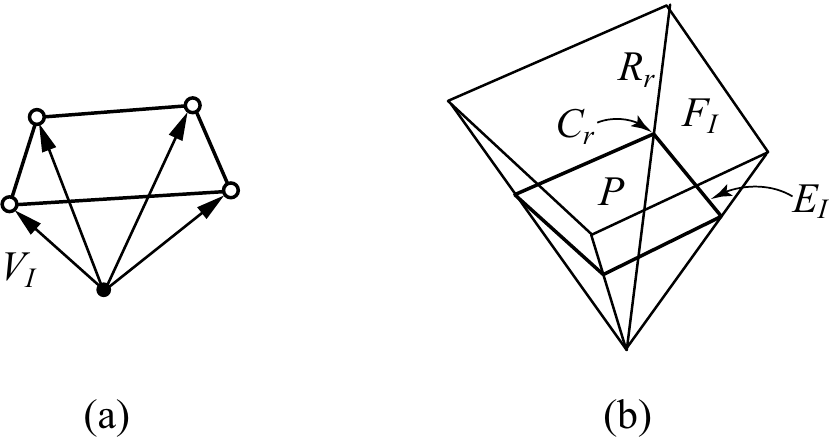}
\caption{(a) A toric diagram. (b) The corresponding base $B$.}\label{fig1.fig}
\end{figure}

The toric diagram of a toric Calabi-Yau three-fold is a convex polygon
drawn in the two-dimensional lattice $L_\ZZ\subset L=\RR^2$.
We label lattice points on the boundary of the diagram
by $I=1,2,...,d$ in counter-clockwise order,
where $d$ is the perimeter of the diagram.
We treat the label $I$ as a cyclic variable with the
relation $I+d\approx I$.
We label edges on the boundary of the toric diagram
by $r\in\ZZ+1/2$, which is also treated as a cyclic variable.
We assign $r=I+1/2$ to the edge between vertices $I$ and $I+1$.
Let $\bm{V}_I$ be the two-dimensional integral vector pointing the vertex $I$ in $L_\ZZ$.
We also define three-dimensional integral vectors
$V_I=(\bm{V}_I,1)$.
We denote the three-dimensional lattice
$\ZZ^3$ with $V_I$
defined in it by ${\cal V}_{\ZZ}\in{\cal V}=\RR^3$.
${\cal V}$ and $\wt{\cal V}$ are dual spaces to each other
and the inner product ${\cal V}\otimes\wt{\cal V}\rightarrow\RR$ is defined.
The polyhedral cone in ${\cal V}$ spanned by $V_I$ are called the toric cone.
(See Figure \ref{fig1.fig} (a).)

The toric data expressed as
the set of vectors $V_I\in{\cal V}_\ZZ$ has two roles.
One is to define the base manifold $B$.
$B$ is defined as the dual cone of the toric cone by
\begin{align}
B=\{y\in\wt{\cal V}|V_I\cdot y\geq 0\quad
\forall I\}.
\end{align}
(See Figure \ref{fig1.fig} (b).)
For each $I$ we define the facet $F_I$ by
\begin{align}
F_I=\{y\in B|V_I\cdot y=0\}.
\label{faceteq}
\end{align}
Let $R_r$ with $r=I+1/2$ be the ridge shared by the two adjacent facets $F_I$ and $F_{I+1}$.
If a vertex $I$ is not a corner but is on the side between two corners $I_1$ and $I_2$ (\ref{faceteq}) gives not
a plane but a line.
We can regard such an $F_I$ as a shrinking facet.

The five-dimensional Sasaki-Einstein manifold $SE_5$
is the subset of the Calabi-Yau cone defined by $\rho=1$.
The radial coordinate $\rho$ in the Calabi-Yau cone is given by $\rho=b\cdot y$,
where $b\in{\cal V}$ is a particular vector inside the toric cone called the Reeb vector.
The cross section of $B$ defined by $b\cdot y=1$ is a polygon $P$.
The Sasaki-Einstein manifold $SE_5$ is given as the $\bm{T}^3$
fibration over $P$.
We denote the edge $F_I\cap P$ and the corner
$R_r\cap P$ by $E_I$ and $C_r$, respectively.

The fiber shrinks in a particular manner on the boundary of $P$.
The second role of the toric data is to specify how the fiber shrinks on the boundary.
At a generic point in $P$ the fiber is $\bm{T}^3$, which is identified with ${\cal V}/{\cal V}_\ZZ$,
and we can specify a cycle in the $\bm{T}^3$ by a vector in ${\cal V}_\ZZ$.
On an edge $E_I$ the cycle $V_I$ shrinks
and at a generic point in $E_I$ the fiber is $\bm{T}^2$.
At a corner $C_r$ two cycles $V_{r\pm1/2}$ shrink
at the same time and the fiber becomes $\bm{S}^1$.
For each edge $E_I$ the fibration over $E_I$ gives
a closed three-dimensional manifold,
which we denote by $S_I$.

If $I_1$ and $I_2$ are two adjacent corners of the toric diagram
and $k:=I_2-I_1>1$ then there are $k-1$ vertices between $I_1$ and $I_2$,
and the corresponding cycles $S_I$ are shrinking
at the degenerate corners $C_{I_1+1/2}=\cdots=C_{I_2-1/2}$.
This means the existence of $A_k$ type singularity along the $\bm{S}^1$ fiber
over the degenerate corners.
Each shrinking three-cycle $S_I$ ($I_1<I<I_2$) is
the direct product of the $\bm{S}^1$ and a shrinking two-cycle at the singularity.

\subsection{Quiver gauge theories}
There is an algorithm \cite{Feng:2000mi,Feng:2002zw} to obtain
from the toric data
the quiver diagram of the gauge theory realized on D3-branes
placed at the apex of a toric manifold.
We will not explain it in detail
but comment only on facts relevant to our analysis.
It is known that the number of the anomaly-free $U(1)$ global internal symmetries
is the same as the perimeter $d$ of the toric diagram.
They include
\begin{itemize}
\item one superconformal $R$-symmetry,
\item two mesonic symmetries, and
\item $d-3$ baryonic symmetries.
\end{itemize}
The mesonic and baryonic symmetries are also called
flavor symmetries.
The $R$-symmetry and the mesonic symmetries
act on the $SE_5$ as isometries.
If the manifold has a non-Abelian isometry group
they are the Cartan part of it.

An $R$-symmetry is defined as a symmetry that acts on the
supercharges in a specific way.
This condition, however, does not fix the $R$-charge,
and there is an ambiguity to mix the other $d-1$ charges that
do not acts on supercharges.
We can determine the $R$-charge appearing in the superconformal
algebra by using $a$-maximization \cite{Intriligator:2003jj}
(or, equivalently, volume minimization \cite{Martelli:2005tp} on the gravity side.)

For analysis of a general toric quiver gauge theory
there is a convenient basis of $d$ independent charges
rather than the basis associated with the classification above.
We denote them by $R_I$ ($I=1,\ldots,d$), and
each of them corresponds to a vertex on the boundary of the toric diagram.
We use a bipartite graph to connect
a toric diagram and a corresponding quiver gauge theory,
and on the gauge theory side $R_I$ are defined with perfect
matchings in the bipartite graph \cite{Franco:2005rj,Butti:2005vn,Butti:2005ps}.
We normalize them so that all bi-fundamental fields carry charges $0$ or $+1$.
All $R_I$ acts on the supercharge in the same manner and hence they are all
$R$-charges with unusual normalization;
The supercharges carry $R_I=\pm 1/2$.
There is a prescription
to associate each perfect matching on the graph with
a vertex in the toric diagram.
For each corner of the toric diagram
there is a unique perfect matching
and we can uniquely define the corresponding $R_I$,
while for other vertices on the boundary
we have more than one perfect matchings
and we need to choose one of them
to define $R_I$.
These $d$ charges form a basis of the global symmetries,
and all charges listed above are linear combination of $R_I$ in the
form
\begin{align}
\sum_{I=1}^dc_IR_I.
\end{align}
For an $R$-charge with the standard normalization
$\sum_{I=1}^dc_I=2$,
while for a flavor symmetry $\sum_{I=1}^dc_I=0$.
An important and convenient property of these charges
is that their geometric action on $SE_5$ is
specified by $V_I$.
Therefore, for a baryonic symmetry,
which has no geometric action,
the coefficients satisfy
$\sum_{I=1}^dc_IV_I=0$.
This is consistent with the fact that
the rank of the baryonic symmetry is $d-3$.

On the gravity side $R_I$ is the angular momentum
associated with the geometric action $V_I$.
A wrapped D3-brane can carry angular momenta even if it stays still
due to the coupling to the background RR $4$-form
potential field $C_4$,
and the values
depend on the gauge choice of $C_4$.
We can specify the gauge choice by specifying the singular locus of the
potential field just like the Dirac string of a Dirac monopole.
The singularity of $C_4$ is expressed as a three-cycle in $SE_5$.
$R_I$ is defined with $C_4$ with the singularity on $S_I$.
With this definition we can show that a D3-brane wrapped over $S_I$
carries
\begin{align}
R_{I'}=N\delta_{II'}.
\label{rid3}
\end{align}

For a vertex $I$ which is not a corner there exist
more than one perfect matchings,
and hence the definition of $R_I$ is ambiguous.
This is related to the ambiguity of choosing a basis
of three-cycles at the singularity.
At $A_{k-1}$ singularity there are $k-1$ shrinking cycles.
We can relate these cycles to simple roots of $A_{k-1}$
algebra, and we have ambiguity in the choice of the simple
roots from the root system.
There are $k!$ bases related by Weyl reflections.
Once we fix a basis of shrinking two-cycles
we can choose $R_I$ so that the charge relation (\ref{rid3}) holds.

On the gauge theory side
the baryonic symmetry is defined as the
anomaly-free subgroup
of the abelian group
defined by replacing all gauge
groups $SU(N)$ by $U(1)$.
It is in general has the form
\begin{align}
U(1)^{d-3}\times G_{\rm disc},
\end{align}
where $G_{\rm disc}$ is a discrete group.
We neglect $G_{\rm disc}$ in the main text
for simplicity,
and give some analysis with $G_{\rm disc}$ in Appendix \ref{refinedb.sec}.

\subsection{Superconformal index}
We define the superconformal index by
\begin{align}
{\cal I}=\tr_{\rm BPS}\left[e^{2\pi i(J+\ol J)}q^{3\ol J}y^{2J}\prod_{I=1}^dv_I^{R_I}\right],
\label{indded}
\end{align}
where $R_I$ are $R$-charges defined in the previous subsection
and $J$ and $\ol J$ are the Lorentz spins.
The trace $\tr_{\rm BPS}$ is the summation over states
(operators) saturating the BPS bound
\begin{align}
\{Q^\dagger,Q\}=H-2\ol J-\frac{3}{2}r_*\geq0,
\end{align}
where $Q$ is the supercharge with the quantum numbers
\begin{align}
[J,Q]=0,\quad
[\ol J,Q]=-\frac{1}{2}Q,\quad
[R_I,Q]=\frac{1}{2}Q.
\end{align}
For (\ref{indded}) to be the superconformal index the fugacities
must satisfy
\begin{align}
\prod_{I=1}^dv_I=q^3.
\end{align}
Notice that to calculate the index (\ref{indded})
we do not have to know the superconformal $U(1)_R$ charge
$r_*$.

In the numerical calculation we necessarily introduce
a maximum order at which we cut off the infinite series.
For this purpose we need to define ``the order'' for each term
in the index as a linear combination of quantum numbers
appearing in the index as exponents of fugacities.
We can use the linear combination $3\ol J+\frac{3}{2}r$
of $\ol J$ and an appropriately chosen $R$-charge $r$.
A choice of $r$ given by
\begin{align}
r=\sum_{I=1}^d c_IR_I,\quad
\sum_{I=1}^dc_I=2
\label{rcharge}
\end{align}
is implemented in the calculation by the variable change
\begin{align}
v_I = u_I q^{\frac{3}{2}c_I}
\end{align}
with new fugacities $u_I$ constrained by
\begin{align}
\prod_{I=1}^d u_I=1.
\end{align}

After this variable change the order agrees with the exponent
of the fugacity $q$.
We can use an arbitrary set of $c_I$ satisfying the constraint in (\ref{rcharge})
as far as the $R$-charge is positive
for all BPS operators.
This requires $c_I\geq 0$, where the equality is allowed only for
$I$ corresponding to the shrinking cycles.

A natural choice for $r$ with clear physical meaning is
the superconformal $U(1)_R$ charge $r_*$.
In that case the order is
$3\ol J+\frac{3}{2}r_*=H+\ol J$ for BPS operators,
and up to the spin $\ol J$ the order is identified with the dimension of operators.
However, they are in general irrational,
and then we need special treatment in the numerical calculation.
To simplify the calculation we adopt $r$ with rational coefficients.

\section{Index from AdS}\label{ads.sec}
\subsection{Kaluza-Klein contributions}
In the large $N$ limit the superconformal index is calculated on the gravity side
as the index of the Kaluza-Klein modes, and given by
\begin{align}
{\cal I}^{\rm KK}=\Pexp i^{\rm KK},
\label{ikk}
\end{align}
where $\Pexp$ is the plethystic exponential defined in
(\ref{pexpdef}) and
$i^{\rm KK}$ is the single-particle index of the Kaluza-Klein modes.
For the simplest example with $SE_5=\bm{S}^5$ corresponding to the ${\cal N}=4$ supersymmetric $U(N)$
Yang-Mills theory
$i^{\rm KK}$ was calculated in \cite{Kinney:2005ej}:
\begin{align}
\frac{v_1}{1-v_1}
+\frac{v_2}{1-v_2}
+\frac{v_3}{1-v_3}.
\label{n4index}
\end{align}
(Precisely this is not the single-particle index
of the ${\cal N}=4$ theory given in \cite{Kinney:2005ej}.
We subtracted the contribution of the diagonal $U(1)$ part of the ${\cal N}=1$ vector multiplet
while the diagonal parts of the three adjoint ${\cal N}=1$ chiral multiplets are left included
because we treat adjoint fields as special bi-fundamental fields, for which
we cannot define the diagonal part.)
In the case of an abelian orbifold $\bm{S}^5/\Gamma$,
the contribution of the gravity multiplet is given by
(\ref{n4index}) again with $v_I$ replaced by appropriate powers of them.
In addition, if the orbifold
has singular loci, we need to include the contribution
of tensor multiplets localized on the loci.
The number of tensor multiplet is $d-3$, and
each tensor multiplet
contributes to $i^{\rm KK}$ by \cite{Nakayama:2005mf}
\begin{align}
\frac{w}{1-w}
\label{tensorikk}
\end{align}
with an appropriate choice of $w$.
From these results it is natural to guess the following formula
for a general toric manifold:
\begin{align}
{\cal I}^{\rm KK}=\Pexp\left(
\sum_{r=1/2}^{d-1/2}\frac{w_r}{1-w_r}\right),
\label{toricKK}
\end{align}
where $r$ is a label for edges of the toric diagram
and $w_r$ are fugacities defined for each edge $r$.
Indeed the formula in the form
(\ref{toricKK}) was derived in
\cite{Eager:2012hx,Agarwal:2013pba},
and the fugacities $w_r$ are defined in the following way.

Let us focus on a ridge $r$ of the dual cone
shared by two adjacent facets $F_I$ and $F_{I+1}$ with $I=r-1/2$.
Let $g_r\in\wt{\cal V}_\ZZ$ be the primitive integer vector
along the ridge.
In general a choice of a vector
in $\wt{\cal V}$ defines a coordinate in ${\cal V}$
through the inner product.
We denote the coordinate defined with $g_r$
by $\varphi_r$.
The coordinate $\varphi_r$ has clear geometric meaning.
By definition $g_r$ is orthogonal to
$V_I$ and $V_{I+1}$.
Therefore, on the corresponding cycles in the $\bm{T}^3$ fiber
it takes a constant value, and it is well defined
even if these cycles shrink on the ridge.
Namely, it is a coordinate parameterizing
the $\bm{S}^1$ fiber on the ridge.
The primitivity of $g_r$ means that the coordinate $\varphi_r$
is normalized so that the period of $\bm{S}^1$ is $1$.
We define fugacities associated with the coordinates by
\begin{align}
w_r=\prod_{I=1}^dv_I^{V_I\cdot g_r},
\label{wrdef}
\end{align}
and these are the fugacities appearing in the formula (\ref{toricKK}).

\subsection{Wrapped D3-brane contributions}
Let us discuss the contribution of D3-branes wrapped on three-cycles.

The analyses of orbifold theories \cite{Nakayama:2005mf,Arai:2019wgv}
show that
D3-branes wrapping on shrinking cycles do not contribute
to the index.
The large $N$ index is reproduced
by taking account of only the Kaluza-Klein
contribution of the gravity
and tensor multiplets.
This fact suggests that we have only to consider
D3-branes wrapped on visible cycles
corresponding to corners of the toric diagram.

In this paper we only focus on D3-branes with single wrapping,
and leave the analysis of branes with multiple wrapping
for future work.
A technical remark is in order.
If there are shrinking cycles the definition
of a basis of the shrinking cycles is ambiguous
and the distinction between single wrapping and multiple wrapping
becomes unclear.
To make the following analysis simple as much as possible
we neglect the wrapping numbers for shrinking cycles by setting
the corresponding fugacity to be $1$.
For more detailed explanation for this point refer to Appendix \ref{refinedb2.sec}
in which we show an example of analysis with keeping such fugacities.

With this remark in mind,
we propose the following formula for the index
of a D3-brane wrapped over a visible cycle $S_I$:
\begin{align}
{\cal I}^{\rm D3}_{S_I}=m_Iv_I^N\Pexp i^{\rm D3}_{S_I},\quad
I\in\{\mbox{corners}\}.
\label{id3formula}
\end{align}
The factor $v_I^N$ is the classical contribution
obtained from (\ref{rid3}).
$i^{\rm D3}_{S_I}$ is the single-particle index of the fluctuations
on the wrapped D3-brane, and $m_I$ is a numerical factor representing the
degeneracy associated with the $U(1)$ holonomy on the wrapped D3-brane.
In the following we explain how we can determine $i^{\rm D3}_{S_I}$ and $m_I$
from the toric data.

\begin{figure}[htb]
\centering
\includegraphics{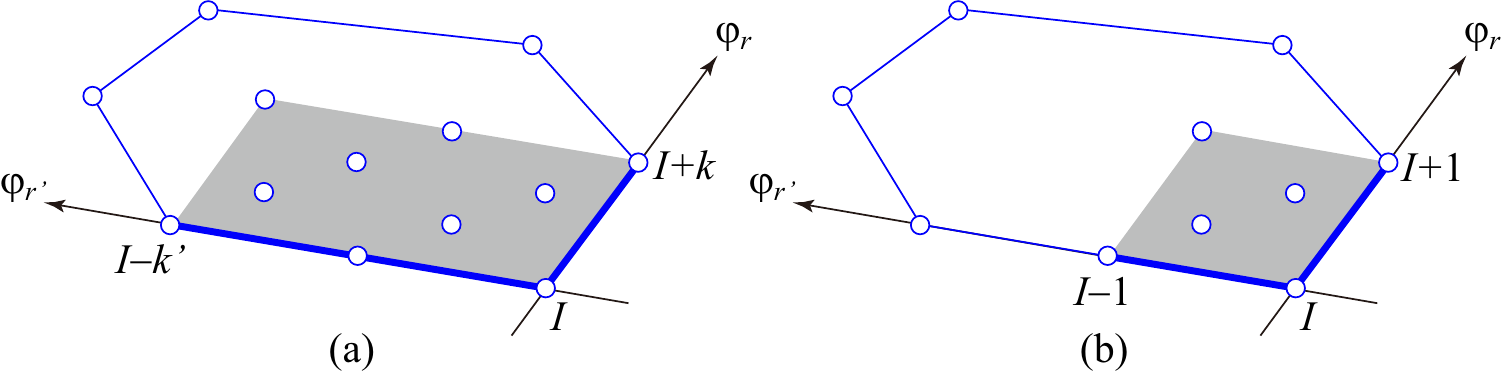}
\caption{Two parallelograms used to describe the
$\bm{T}^2$ fiber over $E_I$.}
\label{ridgecds}
\end{figure}

Let $S_I$ be a visible three-cycle
we are focusing on.
This means $I$ is a corner of the toric diagram.
$S_I$ is a $\bm{T}^2$ fibration over $E_I$ with the fiber shrinks at two ends.
We can identify the $\bm{T}^2$ fiber with $L/L_\ZZ$.
(Remember that $L$ is the plane with the toric diagram drawn on it
and $L_\ZZ$ is the integer lattice in $L$.)
We introduce two coordinates $\varphi_r$ and $\varphi_{r'}$ with
$r=I-1/2$ and $r'=I+1/2$ to parameterize the plane $L$.
The two sides sharing $I$ are given by $\varphi_r=0$ and $\varphi_{r'}=0$.
(See Figure \ref{ridgecds}.)
We regard the plane $L$ as the covering space
of the $\bm{T}^2$ fiber on $E_I$.
The $\bm{T}^2$ fiber
is given as a lattice quotient of the plane
as follows.

Let $I-k'$ and $I+k$ be the corners adjacent with $I$.
$k$ and $k'$ are the lengths of the sides sharing the vertex $I$.
We define $m$ and $n$ by
\begin{align}
m=\varphi_r|_{I+k},\quad
n=\varphi_{r'}|_{I-k'},
\label{mndef1}
\end{align}
and consider the parallelogram on $L$ defined by
(Figure \ref{ridgecds} (a))
\begin{align}
0\leq\varphi_r<m,\quad
0\leq\varphi_{r'}<n.
\end{align}
We regard this as a torus by identifying opposite sides.
If there is no lattice points
in this parallelogram except $\varphi_r=\varphi_{r'}=0$,
this torus is nothing but the fiber at a generic point on $E_I$.
The cycle with constant $\varphi_r$ shrinks at the endpoint $C_r$
while the cycle with constant $\varphi_{r'}$ shrinks at the other endpoint $C_{r'}$.
As the result the three-cycle $S_I$ is topologically $\bm{S}^3$.
If the area of the parallelogram is greater than $1$
we should perform
the orbifolding with $\Gamma$, the discrete group defined by the lattice points
in the parallelogram.
Then the three-cycle $S_I$ is topologically $\bm{S}^3/\Gamma$.

The $U(1)$ holonomy on the D3-brane wrapped on $S_I$ is specified by
an element of the dual group $\wt\Gamma=\Hom(\Gamma,U(1))$,
and the number of elements of $\wt\Gamma$ is the same as $|\Gamma|$,
which is the same as the area of the parallelogram.
Because we consider a single D3-brane no fields on the
brane couple to the $U(1)$ holonomy and existence of $|\Gamma|$ values of the holonomy
affects the index simply as the overall numerical factor.
This gives the degeneracy factor
\begin{align}
m_I=|\Gamma|=(\bm{V}_{I+k'}-\bm{V}_I)\times(\bm{V}_{I-k}-\bm{V}_I).
\label{midef}
\end{align}

We can also take $I-1$ and $I+1$ to define
the parallelogram
(Figure \ref{ridgecds} (b)).
Namely, instead of (\ref{mndef1})
we define $m$ and $n$ by
\begin{align}
m=\varphi_r|_{I+1},\quad
n=\varphi_{r'}|_{I-1}.
\label{mndef2}
\end{align}
If $kk'>1$ this gives a smaller parallelogram.
With this minimal choice we obtain smaller orbifold group $\Gamma'$,
which is related to $\Gamma$ by
\begin{align}
\Gamma'=\Gamma/(\ZZ_k\times\ZZ_{k'}).
\label{gpg}
\end{align}
This should give the same $S_I$ as above.
Namely, the following relation should hold.
\begin{align}
\bm{S}^3/\Gamma=\bm{S}^3/\Gamma'.
\end{align}
This is checked as follows.
From the relation (\ref{gpg})
$\bm{S}^3/\Gamma=(\bm{S}^3/(\ZZ_k\times\ZZ_{k'}))/\Gamma'$
and this is the same as $\bm{S}^3/\Gamma'$
because
$\ZZ_k$ and $\ZZ_{k'}$ act on the coordinates $\varphi_r$ and $\varphi_{r'}$
independently and
$\bm{S}^3/(\ZZ_k\times\ZZ_{k'})$ is topologically the same as $\bm{S}^3$.

Even if we use the expression $S_I=\bm{S}^3/\Gamma'$
the multiplicity $m_I$ is still given by
(\ref{midef}) because
when we use $\Gamma'$ the $\bm{S}^3$ before the orbifolding has
type $A_{k-1}$ singular locus if $k>1$ and type $A_{k'-1}$ singular locus if $k'>1$.
These singular loci cause the additional factors $k$ and $k'$ in the multiplicity factor,
and we again obtain $m_I=kk'|\Gamma'|=|\Gamma|$.

Next let us determine $i^{\rm D3}_{S_I}$, the single-particle index of
the fluctuations of the fields on a D3-brane wrapped on $S_I$.
In the case of $SE_5=\bm{S}^5$
it is \cite{Arai:2019xmp}
\begin{align}
f(q,y,w_r,w_{r'})
&=1-\frac{(1-q^{-3}w_rw_{r'})(1-q^{\frac{3}{2}}y)(1-q^{\frac{3}{2}}y^{-1})}
{(1-w_r)(1-w_{r'})},
\label{thed3index}
\end{align}
where $w_r$ and $w_{r'}$ are fugacities associated with
the two ridges $r=I-1/2$ and $r'=I+1/2$.

We claim that if $S_I$ is topologically $\bm{S}^3$ the formula
(\ref{thed3index})
gives the correct answer even for
a general toric manifold for which $\bm{S}^3$ is not always round.
The reason is as follows.
The index can be regarded as the partition function of the theory defined in $\bm{S}^3\times\bm{S}^1$
with appropriate background fields.
In general, a supersymmetric partition function in a supersymmetric background
depends only on a small number of parameters of the background fields \cite{Closset:2013vra}.
In the case of spacetime with the topology $\bm{S}^3\times\bm{S}^1$ the parameters are nothing but the fugacities,
and even for a deformed $\bm{S}^3$ the index should take the same form as the one for the round $\bm{S}^3$
with appropriately chosen fugacities depending on the background fields.
The fugacities can be determined as parameters for the complex structure of the manifold and the
moduli of the normal bundle following the detailed analysis in \cite{Closset:2013vra}.
Fortunately, in the case of toric theories we are studying here we can
skip this analysis.
In the basis of charges used in the definition (\ref{indded}) of the index
all charges are quantized in units of $1/2$ and quantum numbers of BPS states
do not change under continuous deformations as long as
the deformations respect the supersymmetry and the $U(1)$ symmetries,
and then the index is the same as that for the round sphere.

If $S_I$ is an orbifold $\bm{S}^3/\Gamma$
we need to perform the orbifold projection.
Let $(\alpha_g,\beta_g)$ be the coordinates $(\varphi_r,\varphi_{r'})$ at the lattice point
in the parallelogram corresponding to $g\in\Gamma$.
The single-particle index on $S_I=\bm{S}^3/\Gamma$ is given by
\begin{align}
i^{\rm D3}_{S_I}
&={\cal P}_\Gamma f(q,y,w_r,w_{r'})
\nonumber\\
&\equiv \frac{1}{|\Gamma|}\sum_{g\in\Gamma}
f(q,y,\omega_m^{\alpha_g}w^{\frac{1}{m}}_r,
\omega_n^{\beta_g}w_{r'}^{\frac{1}{n}}
),
\label{fgamma}
\end{align}
where $\omega_k\equiv\exp\frac{2\pi i}{k}$.
This formula reproduces the results
of the orbifold theories in \cite{Arai:2019wgv}.

As a simple consistency check let us confirm that
two descriptions of $S_I$ with the two orbifold groups $\Gamma$ and $\Gamma'$
related by (\ref{gpg}) give the same single-particle index.
We should confirm the relation
\begin{align}
{\cal P}_\Gamma f(q,y,w_r,w_{r'})
={\cal P}_{\Gamma'}f(q,y,w_r,w_{r'}).
\end{align}
Because of the relation (\ref{gpg})
we can decompose ${\cal P}_{\Gamma}$ as ${\cal P}_{\Gamma}={\cal P}_{\Gamma'}{\cal P}_{\ZZ_k\times\ZZ_{k'}}$
where ${\cal P}_{\ZZ_k\times\ZZ_{k'}}$ is defined by
\begin{align}
{\cal P}_{\ZZ_k\times\ZZ_{k'}}f(q,y,w_r,w_{r'})
=\frac{1}{kk'}\sum_{i=0}^{k-1}\sum_{j=0}^{k'-1}f(q,y,\omega_k^iw_r^{\frac{1}{k}},\omega_{k'}^jw_{r'}^{\frac{1}{k'}}).
\end{align}
It is sufficient to show
\begin{align}
f(q,y,w_r,w_{r'})
={\cal P}_{\ZZ_k\times\ZZ_{k'}}f(q,y,w_r,w_{r'}).
\label{ffzz}
\end{align}
By using the explicit form of the function $f$ in (\ref{thed3index})
we can easily confirm (\ref{ffzz}).

\section{Examples}\label{examples.sec}
In this section, we calculate the superconformal index on the gravity side according to
the formula (\ref{id3formula})
and compare the obtained results with those numerically
calculated on the gauge theory side by localization method
summarized in Appendix \ref{localization.sec}
to confirm that our prescription does work correctly.
Due to the limitation of the machine power
we take $N=2$ and $3$.
We will show only the results of $N=2$ case in this section.
We will show the results for $N=3$ in Appendix \ref{n3.sec}.

\subsection{$T^{1,1}$ (conifold)}\label{t11.ssec}
\begin{figure}[htb]
\centering
\includegraphics{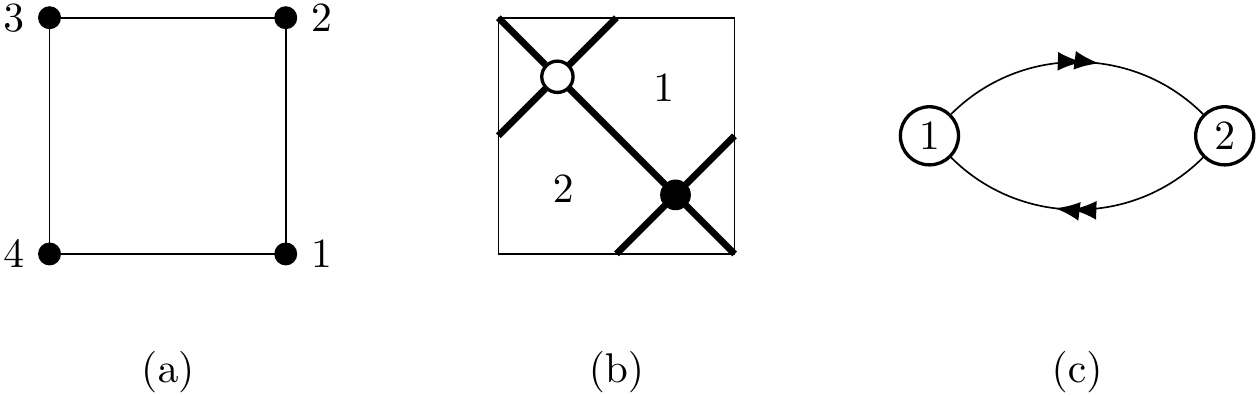}
\caption{The toric diagram of $T^{1,1}$ (a), the corresponding bipartite graph (b) and quiver diagram (c) are shown.}
\label{coniall}
\end{figure}

First we consider $SE_5=T^{1,1}$.
$T^{1,1}$ is the base of the conifold,
and the corresponding boundary theory
is the quiver gauge theory so-called Klebanov-Witten theory \cite{Klebanov:1998hh}.
The toric diagram, the bipartite graph, and the quiver diagram are shown in Figure \ref{coniall}.
The matter contents of this theory is shown in Table \ref{mcconi}.
\begin{table}
\caption{Matter contents and charge assignments are shown for the Klebanov-Witten theory (the conifold).}
\label{mcconi}
\begin{center}
\begin{tabular}{ccccccccc}
\hline
\hline
Fields & $R_1$ & $R_2$ & $R_3$ & $R_4$ & $r_*$ & $F_A$ & $F_B$ & $NB$ \\
\hline
$X_{12}^1(A_1)$ & $1$ & $0$ & $0$ & $0$ & $1/2$ & $1$ & $0$ & $1$ \\
$X_{21}^1(B_1)$ & $0$ & $1$ & $0$ & $0$ & $1/2$ & $0$ & $1$ & $-1$ \\
$X_{12}^2(A_2)$ & $0$ & $0$ & $1$ & $0$ & $1/2$ & $-1$ & $0$ & $1$ \\
$X_{21}^2(B_2)$ & $0$ & $0$ & $0$ & $1$ & $1/2$ & $0$ & $-1$ & $-1$ \\
\hline
\end{tabular}
\end{center}
\end{table}
Corresponding to four vertices of the toric diagram
we have four $R$-charges $R_I$ ($I=1,2,3,4$).
The conifold and $T^{1,1}$ have
the isometry $(SU(2)_A\times SU(2)_B)/\ZZ_2\times U(1)_{r^*}$.
The $U(1)_{r^*}$ factor is the superconformal $R$-symmetry generated by
\begin{align}
\label{scrconi}
r_*=\frac{1}{2}(R_1+R_2+R_3+R_4).
\end{align}
The Cartan generators $F_A$ and $F_B$ of $SU(2)$ factors
are
\begin{align}
F_A=R_1-R_3,\quad
F_B=R_2-R_4.
\label{coniflavor}
\end{align}
The remaining one linear combination of $R_I$
is the baryonic charge
\begin{align}
B=\frac{1}{N}(R_1-R_2+R_3-R_4),
\label{conibaryonic}
\end{align}
which does not act on $T^{1,1}$ geometrically.
On the gauge theory side this can be regarded as the non-diagonal
part of the $U(1)\times U(1)$ symmetry defined by replacing the $SU(N)$ gauge groups by $U(1)$.
Because the theory is non-chiral this symmetry is anomaly-free.
See Table \ref{mcconi}
for charge assignments for these generators.
Corresponding to the charges $F_A$, $F_B$, and $B$,
we introduce the fugacities $u$, $v$, and $\zeta$ by the relation
\begin{align}
v_1^{R_1}
v_2^{R_2}
v_3^{R_3}
v_4^{R_4}
=q^{\frac{3}{2}r_*}
u^{F_A}
v^{F_B}
\zeta^B.
\end{align}
The canonical variables $v_I$ are expressed with new variables by
\begin{align}
v_1=q^{\frac{3}{4}}u\zeta^{\frac{1}{N}},\quad
v_2=\frac{q^{\frac{3}{4}}v}{\zeta^{\frac{1}{N}}},\quad
v_3=\frac{q^{\frac{3}{4}}\zeta^{\frac{1}{N}}}{u},\quad
v_4=\frac{q^{\frac{3}{4}}}{v\zeta^{\frac{1}{N}}}.
\label{T11v1234}
\end{align}
Due to the $SU(2)_A \times SU(2)_B$ flavor symmetry
the index can be written in terms of $SU(2)$ characters
$\chi _n^u=\chi _n(u)$ for $SU(2)_A$ and $\chi _n^v=\chi _n(v)$ for $SU(2)_B$
as well as the $SU(2)$ spin character $\chi _n^J=\chi _n(y)$,
where the functions $\chi_n(c)$ are defined by
\begin{align}
\chi _n(c) =\frac{c^{n+1}-c^{-(n+1)}}{c-c^{-1}}.\label{su2chara}
\end{align}

On the gauge theory side we define
the index for the sector with a specific baryon number $B$ by the expansion
with respect to $\zeta$:
\begin{align}
{\cal I}^{\rm gauge}
=\sum_{B\in\ZZ}{\cal I}_B^{\rm gauge},\quad
{\cal I}_B^{\rm gauge}\propto\zeta ^B.
\end{align}
Instead of defining ${\cal I}_B^{\rm gauge}$ as
the coefficients in the $\zeta$ expansion
we include $\zeta^B$ in ${\cal I}_B^{\rm gauge}$.

On the gravity side the baryonic charge corresponds to
the wrapping number of D3-branes \cite{Gubser:1998fp}.
Remember that $H_3(T^{1,1},\ZZ)=\ZZ$ and
the wrapping sectors are labeled by a single integer $B$,
which is identified with the baryonic charge.
By combining (\ref{rid3}) and (\ref{conibaryonic})
we can determine the wrapping number of each cycle $S_I$.
Equivalently, we can read off the charges as the
exponents of $\zeta$
from the classical
factors
\begin{align}
v_1^N
=q^{\frac{3}{4}N}u^N\zeta,\quad
v_2^N
=\frac{q^{\frac{3}{4}N}v^N}{\zeta },\quad
v_3^N
=\frac{q^{\frac{3}{4}N}\zeta }{u^N},\quad
v_4^N
=\frac{q^{\frac{3}{4}N}}{v^N\zeta }.
\label{icl1234}
\end{align}
By using these we can determine
brane configurations contributing to the index of
a specific wrapping sector and the orders of their contributions.
For example, let us consider $B=1$ sector.
There are two single-wrapping brane configurations
$S_1$ and $S_3$ with $B=1$.
They give ${\cal O}(q^{\frac{3}{4}N})$ contributions
to the index.
Furthermore, we also have multiple-wrapping configurations
like $S_1+S_2+S_3$.
If we naively assume that the contribution is given as the product
of constituent contributions this gives ${\cal O}(q^{\frac{9}{4}N})$ terms.
Expected orders of the corrections obtained in this way
for different wrapping numbers are shown in Table \ref{coni_lowest}.
\begin{table}[htb]
\caption{The lowest and higher order contributions to the indices of the D3-branes with
the wrapping numbers $B\in\ZZ$, and the corresponding brane configurations are shown for $T^{1,1}$.
$S_I+S_J$ denotes the brane configuration  consisting of a brane wrapped on $S_I$ and another brane wrapped on $S_J$.}
\label{coni_lowest}
\vspace{-0.5cm}
\begin{align}
\begin{array}{c|l|l}\hline\hline
 B	&{\rm lowest\ order,\ (config.)} &{\rm higher\ order,\ (config.)} \\\hline
-2	&  & q^{\frac{3N}{2}}, (2S_2,\dots) \\
-1	& q^{\frac{3N}{4}}, (S_2,S_4)	& q^{\frac{9N}{4}}, (S_1+2S_2,\dots)  \\
0	& 1	& q^{\frac{3N}{2}}, (S_1+S_2,\dots) \\
1	& q^{\frac{3N}{4}}, (S_1,S_3) & q^{\frac{9N}{4}}, (2S_1+S_2,\dots)  \\
2	& 	& q^{\frac{3N}{2}}, (2S_1, \dots) \\\hline
\end{array}\nn
\end{align}
\end{table}

Now, let us start the comparison of the results on the
gravity side and those on the gauge theory side.
We consider $N=2$ case here.
See Appendix \ref{n3.sec} for the results for $N=3$.
Let us first consider $B=0$ sector.
The relation expected from the order estimation of
the multiple-wrapping D3-brane contribution is
\begin{align}
\mathcal{I}_0^{\rm gauge}&=\mathcal{I}^{\rm KK}+\mathcal{O}(q^{\frac{3}{2}N}).\label{conirelation_0}
\end{align}
On the gauge theory side
the numerical analysis for $N=2$ gives
\begin{align}
\mathcal{I}_0^{\rm gauge}
&=1+\chi _1^u\chi _1^vq^{\frac{3}{2}}+(2-\chi _2^u-\chi _2^v+2\chi _2^u\chi _2^v)q^3\nn \\
&+(-2\chi _1^u\chi _1^v-\chi _3^u\chi _1^v-\chi _1^u\chi _3^v+2\chi _3^u\chi _3^v+2(1+\chi _2^u\chi _2^v)\chi _1^J)q^{\frac{9}{2}}+\mathcal{O}(q^6).
\label{conigauge_0}
\end{align}
On the gravity side
${\cal I}^{\rm KK}$ was first calculated in \cite{Nakayama:2006ur},
and is given by (\ref{toricKK})
with the fugacities
\begin{align}
&w_{\frac{1}{2}}=v_2v_3=\frac{q^{\frac{3}{2}}v}{u},\quad
w_{1+\frac{1}{2}}=v_3v_4=\frac{q^{\frac{3}{2}}}{uv},\nonumber\\
&w_{2+\frac{1}{2}}=v_4v_1=\frac{q^{\frac{3}{2}}u}{v},\quad
w_{3+\frac{1}{2}}=v_1v_2=q^{\frac{3}{2}}uv.
\label{T11w}
\end{align}
These are invariant under the $\ZZ_2$ action $(u,v)\rightarrow (-u,-v)$.
This is consistent with the fact that
the $\ZZ_2$ quotient group in the isometry group
is generated by $e^{\pi i(F_A+F_B)}$.
Furthermore, these do not contain the baryonic fugacity $\zeta$,
and this is consistent with the fact that the Kaluza-Klein modes
do not have wrapping numbers.
(\ref{toricKK}) gives
\begin{align}
\mathcal{I}^{\rm KK}&=(\cdots \text{identical terms with (\ref{conigauge_0})}\cdots )\nn \\
&+(2\chi _1^u\chi _1^v-\chi _3^u\chi _1^v-\chi _1^u\chi _3^v+3\chi _3^u\chi _3^v)q^{\frac{9}{2}}+\mathcal{O}(q^{6}).
\label{ikkt11}
\end{align}
(\ref{conigauge_0}) and
(\ref{ikkt11}) satisfy the relation (\ref{conirelation_0}).

Next, let us consider $B\neq 0$ sectors.
We have two sectors $B=\pm1$ with the contribution of
single-wrapping branes.
To extract the brane contributions
we consider the ratio ${\cal I}_B/{\cal I}^{\rm KK}$
rather than ${\cal I}_B$.
$S_1$ and $S_3$ carry $B=1$ and $S_2$ and $S_4$ carry $B=-1$.
Expected relations are
\begin{align}
\frac{\mathcal{I}_1^{\rm gauge}}{\mathcal{I}^{\rm KK}}
&=(\mathcal{I}_{S_1}^{\rm D3}+\mathcal{I}_{S_3}^{\rm D3})+\mathcal{O}(q^{\frac{9}{4}N}),\nn \\
\frac{\mathcal{I}_{-1}^{\rm gauge}}{\mathcal{I}^{\rm KK}}
&=(\mathcal{I}_{S_2}^{\rm D3}+\mathcal{I}_{S_4}^{\rm D3})+\mathcal{O}(q^{\frac{9}{4}N}).\label{conirelations}
\end{align}
These two relations are not independent due to a symmetry.
The toric diagram (Figure \ref{coniall} (a)) and the bipartite graph (Figure \ref{coniall} (b)) have the
$\ZZ_4$ rotational symmetry.
The counter-clockwise $\pi/2$ rotation
maps the vertex $I$ to $I+1$ and
the charge $R_I$ to $R_{I+1}$.
The charges $r_*$, $F_A$, $F_B$, and $B$ are mapped as
\begin{align}
r_*\rightarrow r_*,\quad
F_A\rightarrow F_B,\quad
F_B\rightarrow -F_A,\quad
B\rightarrow-B.
\end{align}
Therefore, the index is invariant under
\begin{align}
(q,y,u,v,\zeta )\rightarrow \left (q,y,v,\tfrac{1}{u},\tfrac{1}{\zeta }\right ),
\label{conifugarelation}
\end{align}
and this transforms two relations in (\ref{conirelations}) to each other.
We focus on the $B=1$ sector.
The calculation on the gauge theory side with $N=2$ gives
\begin{align}
\frac{\mathcal{I}_1^{\rm gauge}}{\mathcal{I}^{\rm KK}}&=\zeta \Big [\chi _2^uq^{\frac{3}{2}}+(-\chi _1^u\chi _1^v+(-1+\chi _2^u)\chi _1^J)q^3\nn \\
&+(1-\chi _2^u-\chi _2^v-\chi _2^u\chi _2^v+\chi _1^u\chi _1^v\chi _1^J+(-1+\chi _2^u)\chi _2^J)q^{\frac{9}{2}}\nn \\
&+(2\chi _1^u\chi _1^v-\chi _3^u\chi _1^v+\chi _1^u\chi _3^v-\chi _5^u\chi _3^v\nn \\
&+(1-\chi _2^u-3\chi _2^v+\chi _4^u\chi _2^v)\chi _1^J+(-1+\chi _2^u)\chi _3^J)q^6\nn \\
&+(9+3\chi _2^u+3\chi _4^u-\chi _6^u+\chi _2^v+\chi _4^v\nn \\
&+7\chi _2^u\chi _2^v+2\chi _4^u\chi _2^v+\chi _6^u\chi _2^v+\chi _4^u\chi _4^v-\chi _6^u\chi _4^v\nn \\
&+(-7\chi _1^u\chi _1^v-8\chi _3^u\chi _1^v-2\chi _1^u\chi _3^v-2\chi _3^u\chi _3^v-\chi _5^u\chi _3^v)\chi _1^J\nn \\
&+(1+3\chi _2^u-\chi _4^u+2\chi _2^v+\chi _2^u\chi _2^v+3\chi _4^u\chi _2^v)\chi _2^J+(-1+\chi _2^u)\chi _4^J)q^{\frac{15}{2}}+\mathcal{O}(q^{\frac{17}{2}})\Big ].\label{conigauge_1}
\end{align}
As is mentioned above this is written in terms of $SU(2)$ characters $\chi_n^u$ and $\chi_n^v$.
On the gravity side there are two contributions:
\begin{align}
{\cal I}^{\rm D3}_{S_1}=v_1^N\Pexp f(q,y,w_{\frac{1}{2}},w_{1+\frac{1}{2}}),\quad
{\cal I}^{\rm D3}_{S_3}=v_3^N\Pexp f(q,y,w_{2+\frac{1}{2}},w_{3+\frac{1}{2}}).
\end{align}
Although each of them does not respect the $SU(2)_A$ flavor symmetry,
the sum of two contributions becomes a linear combination
of $SU(2)_A$ characters
by the mechanism explained in detail in \cite{Arai:2019xmp}.
The result for $N=2$ is
\begin{align}
\mathcal{I}^{\rm D3}_{S_1}+\mathcal{I}^{\rm D3}_{S_3}&=\zeta \Big [(\cdots \text{terms identical with (\ref{conigauge_1})}\cdots )\nn \\
&+(\chi _4^u-\chi _6^u+\chi _2^v+2\chi _2^u\chi _2^v-\chi _6^u\chi _2^v+\chi _2^u\chi _4^v-\chi _8^u\chi _4^v\nn \\
&+(-\chi _1^u\chi _1^v-\chi _3^v\chi _1^v+\chi _5^u\chi _1^v-\chi _1^u\chi _3^v-\chi _3^u\chi _3^v+\chi _7^u\chi _3^v)\chi _1^J\nn \\
&+(1-\chi _4^u-\chi _2^v+\chi _2^u\chi _2^v)\chi _2^J+(-1+\chi _2^u)\chi _4^J)q^{\frac{15}{2}}+
\mathcal{O}(q^{\frac{17}{2}})\Big ].
\end{align}
Again, the result is consistent
with the expected relation (\ref{conirelations}).

\subsection{$T^{2,2}$ (complex cone over $F_0$)}\label{f0.sec}
\begin{figure}[htb]
\begin{center}
\includegraphics{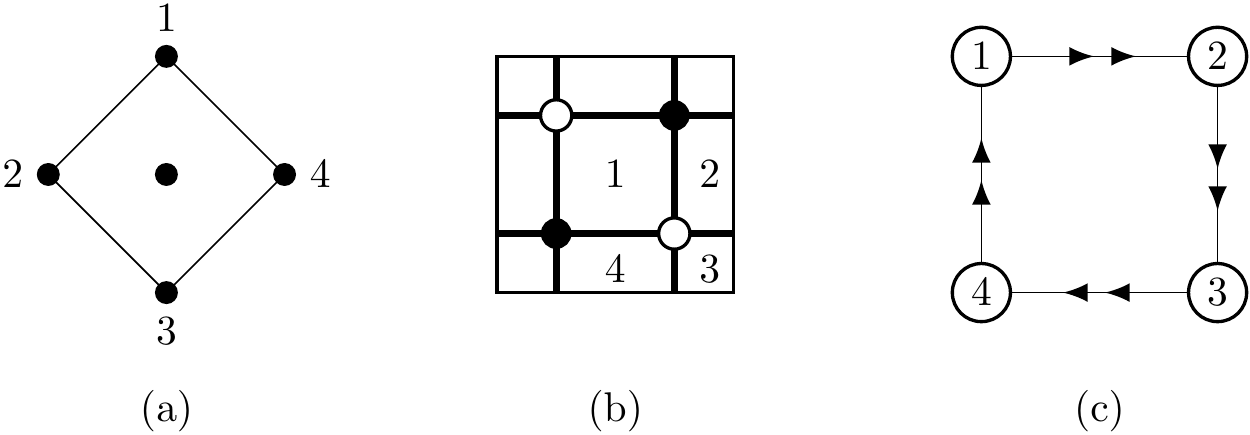}
\caption{The toric diagram of $T^{2,2}$ (a), and the corresponding bipartite graph (b), the corresponding quiver diagram (c) are shown.}
\label{F0toric}
\end{center}
\end{figure}

The next example we discuss is $SE_5=T^{2,2}$.
As is obvious from the toric diagram shown in Figure \ref{F0toric}
this is a $\ZZ_2$ orbifold of $T^{1,1}$.
The corresponding Calabi-Yau cone is a $\ZZ_2$ orbifold of the conifold,
which is often referred to as the complex cone over $F_0$.
The isometry group is $(SU(2)_A/\ZZ_2)\times(SU(2)_B/\ZZ_2)\times U(1)_{r_*}$.
Again, we have four $R$-charges $R_I$ ($I=1,2,3,4$) corresponding to the corners
of the toric diagram.
We define charges $r_*$, $F_1$, $F_2$, and $B$ in the same way as
in the conifold case:
\begin{align}
&r_*=\frac{1}{2}(R_1+R_2+R_3+R_4),\nonumber\\
&F_A=R_1-R_3,\quad
F_B=R_2-R_4,\quad
B=\frac{1}{N}(R_1-R_2+R_3-R_4).
\label{f0charges}
\end{align}
The charge assignments of these charges are shown in Table \ref{F0_charge}.
\begin{table}
\caption{Matter contents and charge assignments are shown for the $F_0$ model.}
\label{F0_charge}
\begin{center}
\begin{tabular}{ccccccccc}
\hline
\hline
Fields & $R_1$ & $R_2$ & $R_3$ & $R_4$ & $r_*$ & $F_A$ & $F_B$ & $NB$  \\
\hline
$X_{12}^1$ & $1$ & $0$ & $0$ & $0$ & $1/2$ & $1$ & $0$ & $1$  \\
$X_{12}^2$ & $0$ & $0$ & $1$ & $0$ & $1/2$ & $-1$ & $0$ & $1$  \\
$X_{23}^1$ & $0$ & $1$ & $0$ & $0$ & $1/2$ & $0$ & $1$ & $-1$  \\
$X_{23}^2$ & $0$ & $0$ & $0$ & $1$ & $1/2$ & $0$ & $-1$ & $-1$  \\
$X_{34}^1$ & $0$ & $0$ & $1$ & $0$ & $1/2$ & $-1$ & $0$ & $1$   \\
$X_{34}^2$ & $1$ & $0$ & $0$ & $0$ & $1/2$ & $1$ & $0$ & $1$  \\
$X_{41}^1$ & $0$ & $0$ & $0$ & $1$ & $1/2$ & $0$ & $-1$ & $-1$ \\
$X_{41}^2$ & $0$ & $1$ & $0$ & $0$ & $1/2$ & $0$ & $1$ & $-1$ \\
\hline
\end{tabular}
\end{center}
\end{table}
Corresponding to the charges in
(\ref{f0charges}) we define the fugacities $u$, $v$, and $\zeta$ by
\begin{align}
v_1^{R_1}
v_2^{R_2}
v_3^{R_3}
v_4^{R_4}
=q^{\frac{3}{2}r_*}
u^{F_A}
v^{F_B}
\zeta^B.
\label{f0fug}
\end{align}
The canonical variables $v_I$ expressed in terms of new ones are
\begin{align}
v_1=q^{\frac{3}{4}}u\zeta^{\frac{1}{N}},\quad
v_2=\frac{q^{\frac{3}{4}}v}{\zeta^{\frac{1}{N}}},\quad
v_3=\frac{q^{\frac{3}{4}}\zeta^{\frac{1}{N}}}{u},\quad
v_4=\frac{q^{\frac{3}{4}}}{v\zeta^{\frac{1}{N}}}.
\end{align}
These relations are identical with those in (\ref{T11v1234}).
The ridge fugacities are given by
\begin{align}
w_{\frac{1}{2}}=v_2^2v_3^2,\quad
w_{1+\frac{1}{2}}=v_3^2v_4^2,\quad
w_{2+\frac{1}{2}}=v_4^2v_1^2,\quad
w_{3+\frac{1}{2}}=v_1^2v_2^2.
\label{f0ridge}
\end{align}
These are square of the corresponding variables
in (\ref{T11w}).
This is a reflection of the fact that $T^{2,2}$ is
the $\ZZ_2$ quotient of $T^{1,1}$.
These are invariant under
two $\ZZ_2$ actions $u\rightarrow-u$
and $v\rightarrow-v$
corresponding to the quotient group
$\ZZ_2^2$ in the isometry group.

According to the general rule there is one continuous baryonic $U(1)$ symmetry.
In addition we have non-trivial discrete factor $G_{\rm disc}=\ZZ_2\times\ZZ_2$.
In this section we neglect this.
See Appendix \ref{refinedb1.sec} for some analysis
with these discrete factors.
Again, we define the index ${\cal I}_B^{\rm gauge}$ of the sector
with a specific baryonic charge $B$ by the expansion
\begin{align}
{\cal I}^{\rm gauge} = \sum_{B\in\ZZ} {\cal I}^{\rm gauge}_B,\quad
{\cal I}_B^{\rm gauge}\propto\zeta^B.
\end{align}

For small values of $B$ we show in Table \ref{f0_lowest}
the expected order of
the corrections due to wrapped branes.
\begin{table}[htb]
\caption{The lowest and higher order contributions to the indices of the D3-branes with
the wrapping numbers $B\in\ZZ$, and the corresponding brane configurations are shown for $T^{2,2}$. }
\label{f0_lowest}
\vspace{-0.5cm}
\begin{align}
\begin{array}{c|l|l}\hline\hline
 B	& {\rm lowest\ order,\ (config.)} &{\rm higher\ order,\ (config.)} \\\hline
-2	&  & q^{\frac{3}{2}N}, (2S_2,\dots) \\
-1	& q^{\frac{3}{4}N}, (S_2,S_4)	& q^{\frac{9}{4}N}, (S_1+2S_2,\dots)  \\
0	& 1	& q^{\frac{3}{2}N}, (S_1+S_2,\dots) \\
1	& q^{\frac{3}{4}N}, (S_1,S_3) & q^{\frac{9}{4}N}, (2S_1+S_2,\dots)  \\
2	& 	& q^{\frac{3}{2}N}, (2S_1, \dots) \\\hline
\end{array}\nn
\end{align}
\end{table}
We focus on the sectors with $B=0$ and $B=\pm1$.
The expected relations are as follows:
\begin{align}
\mathcal{I}_{0}^{\mathrm{gauge}}&=\mathcal{I^\mathrm{KK}}+\mathcal{O}(q^{\frac{3}{2}N}),\nn\\
\frac{\mathcal{I}_{-1}^{\mathrm{gauge}}}{\mathcal{I^\mathrm{KK}}}
&=\mathcal{I}^{\mathrm{D3}}_{S_2}+\mathcal{I}^{\mathrm{D3}}_{S_4}+\mathcal{O}(q^{\frac{9}{4}N}),\nn\\
\frac{\mathcal{I}_{1}^{\mathrm{gauge}}}{\mathcal{I^\mathrm{KK}}}&=\mathcal{I}^{\mathrm{D3}}_{S_1}+\mathcal{I}^{\mathrm{D3}}_{S_3}+\mathcal{O}(q^{\frac{9}{4}N}).
\label{f0relations}
\end{align}
We check these relations for the $N=2$ case.
See also Appendix \ref{n3.sec} for the analysis with $N=3$.
We use
the $SU(2)$ characters
$\chi_n^u$, $\chi_n^v$, and $\chi_n^J$ defined
in Section \ref{t11.ssec}
to write down indices.

On the gauge theory side the localization formula
gives for $B=0$
\begin{align}
\label{f0gauge_0}
\mathcal{I}_{0}^{\mathrm{gauge}}
&=1
+ \left(5 \chi ^u_2 \chi ^v_2-\chi ^u_2-\chi ^v_2+5\right)q^3
+\mathcal{O}(q^{\frac{9}{2}}).
\end{align}
On the gravity side,
(\ref{toricKK}) with the fugacities
(\ref{f0ridge})
gives
\begin{align}
\mathcal{I}^{\mathrm{KK}}&=1+\left(\chi ^u_2 \chi ^v_2-\chi ^u_2-\chi ^v_2+1\right)q^3+\mathcal{O}(q^{6}).
\label{f0ikk}
\end{align}
Comparing (\ref{f0gauge_0}) and (\ref{f0ikk})
we find the first relation in (\ref{f0relations}) holds.

Next, let us consider the sectors with $B=\pm1$.
As in the conifold case
the toric diagram and the bipartite graph
have the $\ZZ_4$ rotational symmetry and
$B=+1$ sector and $B=-1$ sector are related by
the variable change (\ref{conifugarelation}).
We consider only the $B=+1$ sector.
On the gauge theory side we obtain
\begin{align}
\frac{\mathcal{I}_{1}^{\mathrm{gauge}}}{\mathcal{I}^{\mathrm{KK}}}
&=\zeta\Big[
2  \chi ^u_2q^{\frac{3}{2}}+2\left(\chi ^u_2-1\right) \chi ^J_1q^3\nn\\
&+2\left(3 \chi ^u_4\chi ^v_2+\chi ^v_2+2\chi ^u_2-1+\left(\chi ^u_2-1\right)\chi ^J_2\right)q^{\frac{9}{2}}
+\mathcal{O}\left(q^{6}\right)\Big].
\label{f0gauge_1}
\end{align}
On the gravity side
the index for each supersymmetric cycle is given by
\begin{align}
\mathcal{I}^{\mathrm{D3}}_{S_I}=2v_I^N\Pexp\left[\frac{
f(q,y,w_{I-\frac{1}{2}}^{\frac{1}{2}},w_{I+\frac{1}{2}}^{\frac{1}{2}})
+f(q,y,-w_{I-\frac{1}{2}}^{\frac{1}{2}},-w_{I+\frac{1}{2}}^{\frac{1}{2}})
}{2}\right].
\end{align}
The sum of two contributions ${\cal I}_{S_1}^{\rm D3}$ and ${\cal I}_{S_3}^{\rm D3}$  for the $N=2$ case is
\begin{align}
\mathcal{I}^{\mathrm{D3}}_{S_1}+\mathcal{I}^{\mathrm{D3}}_{S_3}&=\zeta\Big[(\cdots \mathrm{identical\ terms\ with}\ (\ref{f0gauge_1})\cdots)\nn\\
   &+ 2  \left(-\chi ^u_2-2 \chi ^v_2-1+\left(\chi ^u_2 -1\right)\chi ^J_2\right)q^{\frac{9}{2}}+\mathcal{O}\left(q^{6}\right)\Big].
\end{align}
We find the last relation in (\ref{f0relations}) holds.

\subsection{$Y^{2,1}$ (complex cone over $dP_1$)}
\begin{figure}[htb]
\begin{center}
\includegraphics{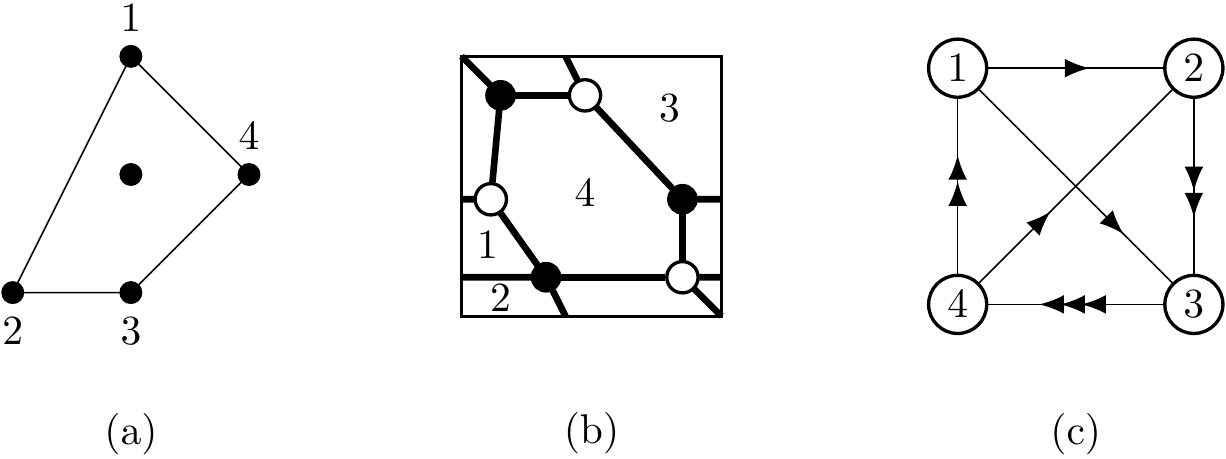}
\caption{The toric diagram of $Y^{2,1}$ (a), the corresponding bipartite graph (b), and the corresponding quiver diagram (c) are shown.}
\label{y21all.eps}
\end{center}
\end{figure}

There are a family of $SE_5$ denoted by $Y^{p,q}$.
We consider $Y^{2,1}$ as the simplest example.
The corresponding Calabi-Yau cone is
the complex cone over the first del Pezzo surface ($dP_1$).
The toric diagram, the bipartite graph, and the quiver diagram
are shown in figure \ref{y21all.eps}.

Generally, the isometry of $Y^{p,q}$ is $SU(2)\times U(1) \times U(1)$,
and the toric diagram has $d=4$.
For $Y^{2,1}$
we define generators of the flavor symmetry by
\begin{align}
F_1=R_2-R_4,\quad F_2=\frac{1}{2}(R_1-R_3),\quad B=\frac{1}{N}\left(2(R_2+R_4)-R_1-3R_3\right).
\label{y21charges}
\end{align}
$F_1$ is the Cartan part of the $SU(2)$ isometry and $F_2$ generates one of the $U(1)$ isometries.
For $U(1)_R$ charge we take
\begin{align}
r=\frac{1}{2}(R_1+R_2+R_3+R_4)
\end{align}
for simplicity of the index calculation
rather than the one in the superconformal algebra \cite{Martelli:2004wu}:
\begin{align}
r_*=(-3+\sqrt{13})R_1
+\frac{16-4\sqrt{13}}{3}(R_2+R_4)
+\frac{-17+5\sqrt{13}}{3}R_3.
\end{align}
The charge assignments are given in Table \ref{Y21_charge}.
\begin{table}
\caption{Matter contents and charge assignments are shown for the $dP_1$ model.}\label{Y21_charge}
\begin{center}
\begin{tabular}{ccccccccc}
\hline
\hline
Fields & $R_1$ & $R_2$ & $R_3$ & $R_4$ & $r$ & $F_1$ & $F_2$ & $NB$ \\
\hline
$X_{12}$ & $0$ & $0$ & $1$ & $0$ & $1/2$ & $0$ & $-1/2$ & $-3$ \\
$X_{23}^1$ & $0$ & $1$ & $0$ & $0$ & $1/2$ & $1$ & $0$ & $2$ \\
$X_{23}^2$ & $0$ & $0$ & $0$ & $1$ & $1/2$ & $-1$ & $0$ & $2$ \\
$X_{34}^1$ & $0$ & $1$ & $1$ & $0$ & $1$ & $1$ & $-1/2$ & $-1$ \\
$X_{34}^2$ & $0$ & $0$ & $1$ & $1$ & $1$ & $-1$ & $-1/2$ & $-1$ \\
$X_{34}^3$ & $1$ & $0$ & $0$ & $0$ & $1/2$ & $0$ & $1/2$ & $-1$ \\
$X_{41}^1$ & $0$ & $1$ & $0$ & $0$ & $1/2$ & $1$ & $0$ & $2$ \\
$X_{41}^2$ & $0$ & $0$ & $0$ & $1$ & $1/2$ & $-1$ & $0$ & $2$ \\
$X_{13}$ & $1$ & $0$ & $0$ & $0$ & $1/2$ & $0$ & $1/2$ & $-1$ \\
$X_{42}$ & $1$ & $0$ & $0$ & $0$ & $1/2$ & $0$ & $1/2$ & $-1$ \\
\hline
\end{tabular}
\end{center}
\end{table}
We introduce new fugacities $u$, $v$, and $\zeta$ by
\begin{align}
v_1^{R_1}v_2^{R_2}v_3^{R_3}v_4^{R_4}
=q^{\frac{3}{2}r}u^{F_1}v^{F_2}\zeta^B.
\end{align}
The fugacities $v_I$ in terms of new ones are
\begin{align}
v_1=\frac{q^{\frac{3}{4}}v^{\frac{1}{2}}}{ \zeta^{\frac{1}{N}}},\quad v_2=q^{\frac{3}{4}}u \zeta^{\frac{2}{N}},\quad
v_3=\frac{q^{\frac{3}{4}}}{v^{\frac{1}{2}} \zeta^{\frac{3}{N}}},\quad v_4=\frac{q^{\frac{3}{4}} \zeta^{\frac{2}{N}}}{u}.
\end{align}

We have one
continuous baryonic $U(1)$ symmetry.
We can easily confirm on the gauge theory side
that it is the full
anomaly-free baryonic symmetry and there is no discrete factor
in this example.
On the gauge theory side
we define the index ${\cal I}_B^{\rm gauge}$ of the sector
with a specific baryonic charge $B$ by the expansion
\begin{align}
{\cal I}^{\rm gauge} = \sum_{B\in\ZZ} {\cal I}^{\rm gauge}_B,\quad
{\cal I}_B^{\rm gauge}\propto\zeta^B.
\end{align}

On the gravity side
the expected orders of corrections due to wrapped D3-branes
are shown in Table \ref{y21_lowest}.
\begin{table}[htb]
\caption{The lowest and higher order contributions to the indices of the D3-branes with
the wrapping numbers $B\in\ZZ$, and the corresponding brane configurations are shown for $Y^{2,1}$.}
\label{y21_lowest}
\vspace{-0.5cm}
\begin{align}
\begin{array}{c|l|l}\hline\hline
 B	&{\rm lowest\ order,\ (config.)}   &{\rm higher\ order,\ (config.)} \\\hline
-3	& q^{\frac{3}{4}N}, (S_3) &  q^{\frac{9}{4}N}, (3S_1) \\
-2	&  & q^{\frac{3}{2}N}, (2S_1)	\\
-1	& q^{\frac{3}{4}N}, (S_1)	& q^{\frac{3}{2}N}, (S_2+S_3, \dots) \\
0	& 1 & q^{\frac{9}{4}N}, (2S_1+S_2) \\
1	&  & q^{\frac{3}{2}N} ,(S_1+S_2, \dots) \\
2	& q^{\frac{3}{4}N}, (S_2,S_4) & q^{3N}, (2S_1+2S_2, \dots) \\
3	&  	&q^{\frac{9}{4}N}, (S_1+2S_2, \dots) \\\hline
\end{array}\nn
\end{align}
\end{table}
There are three sectors $B=-1,-3$, and $2$
that receive corrections from
single-wrapping brane configurations.
We focus on these three sectors and the $B=0$ sector.
The expected relations for these sectors are
\begin{align}
\mathcal{I}_{0}^{\mathrm{gauge}}&=\mathcal{I^\mathrm{KK}}+\mathcal{O}(q^{\frac{9}{4}N}),\nn\\
\frac{\mathcal{I}_{-1}^{\mathrm{gauge}}}{\mathcal{I^\mathrm{KK}}}
&= \mathcal{I}^{\mathrm{D3}}_{S_1}+\mathcal{O}(q^{\frac{3}{2}N}),\nn\\
\frac{\mathcal{I}_{-3}^{\mathrm{gauge}}}{\mathcal{I^\mathrm{KK}}}&=\mathcal{I}^{\mathrm{D3}}_{S_3}+\mathcal{O}(q^{\frac{9}{4}N}),\nn\\
\frac{\mathcal{I}_{2}^{\mathrm{gauge}}}{\mathcal{I^\mathrm{KK}}}&=\mathcal{I}^{\mathrm{D3}}_{S_2}+\mathcal{I}^{\mathrm{D3}}_{S_4}+\mathcal{O}(q^{3N}).\label{y21relations}
\end{align}
We check these relations for $N=2$.
See also Appendix \ref{n3.sec} for the analysis with $N=3$.

For the $B=0$ sector the localization formula with $N=2$ gives
\begin{align}
\label{y21gauge_0}
\mathcal{I}_{0}^{\mathrm{gauge}}&=1+ v \chi ^u_1q^{\frac{9}{4}}+ \left(\frac{\chi ^u_3}{v}-\frac{\chi ^u_1}{v}\right)q^{\frac{15}{4}}+\left(10 v^2 \chi ^u_2-v^2\right)q^{\frac{9}{2}}
+\mathcal{O}(q^{\frac{21}{4}}).
\end{align}
We use $SU(2)$ characters $\chi_n^u=\chi_n(u)$ and $\chi_n^J=\chi_n(y)$.
On the gravity side the contribution of the Kaluza-Klein
modes
is given by the formula (\ref{toricKK}) with the fugacities
\begin{align}
w_{\frac{1}{2}}=v_2^3v_3^2,\quad
w_{1+\frac{1}{2}}=v_3^2v_4^3,\quad
w_{2+\frac{1}{2}}=v_1^2v_4,\quad
w_{3+\frac{1}{2}}=v_1^2v_2.
\end{align}
The result is
\begin{align}
\label{y21kk}
\mathcal{I}^{\mathrm{KK}}&=1+v \chi ^u_1q^{\frac{9}{4}}+\left(\frac{\chi ^u_3}{v}-\frac{\chi ^u_1}{v}\right)q^{\frac{15}{4}}+\left(2v^2 \chi ^u_2-v^2\right)q^{\frac{9}{2}}
+\mathcal{O}(q^{6}),
\end{align}
and we find the first relation in (\ref{y21relations}).

For the $B=-1$ sector, the localization formula with $N=2$ gives
\begin{align}
\label{y21gauge_-1}
\frac{\mathcal{I}_{-1}^{\mathrm{gauge}}}{\mathcal{I}^{\mathrm{KK}}}&=\frac{1}{\zeta}\left[3 vq^{\frac{3}{2}} +3\chi ^u_1q^{\frac{9}{4}}+\left(\frac{3 \chi ^u_2}{v}+3 v \chi ^J_1\right)q^3+0q^{\frac{15}{4}}
+\mathcal{O}(q^{\frac{9}{2}})\right].
\end{align}
On the gravity side the corresponding brane contribution is
\begin{align}
\mathcal{I}^{\mathrm{D3}}_{S_1}
&=3v_1^N\Pexp\left(\frac{1}{3}\sum_{k=0}^2f(q,y,\omega_3^kw_{\frac{1}{2}}^{\frac{1}{3}},\omega_3^kw_{1+\frac{1}{2}}^{\frac{1}{3}})\right)\nonumber\\
&\stackrel{(N=2)}{=}
\frac{1}{\zeta}\left[(\cdots \mathrm{identical\ terms\ with}\
(\ref{y21gauge_-1})\cdots)
+3\frac{ \chi ^u_3}{v^2}q^{\frac{15}{4}}+\mathcal{O}(q^{\frac{9}{2}})\right].
\end{align}
These relations are consistent with the second relation in (\ref{y21relations}).

For the $B=-3$ sector, the localization formula
with $N=2$ gives
\begin{align}
\label{y21gauge_-3}
\frac{\mathcal{I}_{-3}^{\mathrm{gauge}}}{\mathcal{I}^{\mathrm{KK}}}&=\frac{1}{\zeta^{3}}\left[\frac{q^{\frac{3}{2}}}{v}
+ \left(v+\frac{\chi ^J_1}{v}\right)q^3-\chi ^u_1q^{\frac{15}{4}}
+\left(10v^3-\frac{1}{v}+\frac{\chi ^J_2}{v}\right)q^{\frac{9}{2}} +\mathcal{O}(q^{\frac{21}{4}})\right].
\end{align}
On the gravity side we obtain
\begin{align}
\mathcal{I}^{\mathrm{D3}}_{S_3}
&=v_3^N\Pexp\left(f(q,y,w_{2+\frac{1}{2}},w_{3+\frac{1}{2}})\right)\nn\\
&\stackrel{(N=2)}{=}\frac{1}{\zeta^{3}}\left[(\cdots \mathrm{identical\ terms\ with}\ (\ref{y21gauge_-3})\cdots)
+\left(v^3-\frac{1}{v}+\frac{\chi ^J_2}{v}\right)q^{\frac{9}{2}} +\mathcal{O}(q^{6})\right].
\end{align}
We see the third relation on (\ref{y21relations}) holds.

For the $B=2$ sector, the gauge theory result is
\begin{align}
\label{y21gauge_2}
\frac{\mathcal{I}_{2}^{\mathrm{gauge}}}{\mathcal{I}^{\mathrm{KK}}}
&=\zeta^{2}\Big[ 2 \chi ^u_2q^{\frac{3}{2}}+2\left(\chi ^u_2-1\right) \chi ^J_1q^3 -2v \chi
   ^u_1q^{\frac{15}{4}}\nn\\
   &+2\left(- \chi ^u_2-3+\left( \chi ^u_2-1\right) \chi ^J_2\right)q^{\frac{9}{2}}
   +2\left( -\frac{\chi ^u_1}{v}+\chi ^u_1 v \chi ^J_1\right)q^{\frac{21}{4}}\nn\\
   &+ 2\left(
   - v^2\chi ^u_2 +5v^2\chi ^u_4 +4v^2 +\left(\chi ^u_2-1\right) \chi ^J_3\right)q^6+\mathcal{O}(q^{\frac{27}{4}})\Big].
\end{align}
On the gravity side the contribution of $S_2$ and $S_4$ are
\begin{align}
\mathcal{I}^{\mathrm{D3}}_{S_2}
&=2v_2^N
\Pexp\left(\frac{\sum_{\pm}f(q,y,\pm w_{1+\frac{1}{2}}^{\frac{1}{2}},\pm w_{2+\frac{1}{2}}^{\frac{1}{2}})}{2}\right),\\
\mathcal{I}^{\mathrm{D3}}_{S_4}
&=2v_4^N
\Pexp\left(\frac{\sum_{\pm}f(q,y,\pm w_{3+\frac{1}{2}}^{\frac{1}{2}},\pm w_{4+\frac{1}{2}}^{\frac{1}{2}})}{2}\right).
\end{align}
These are summed up to
\begin{align}
\mathcal{I}^{\mathrm{D3}}_{S_2}+\mathcal{I}^{\mathrm{D3}}_{S_4}
&=2\zeta^2\chi_N(u)q^{\frac{3}{4}N}+\cdots
\nonumber\\
&\stackrel{(N=2)}{=}\zeta^2\Big[(\cdots \mathrm{identical\ terms\ with}\ (\ref{y21gauge_2})\cdots)\nn\\
&+2 \left(-\chi ^u_2v^2-v^2+\left(\chi ^u_2  -1\right)\chi ^J_3\right)q^6 +\mathcal{O}(q^{\frac{27}{4}})\Big].
\label{is2s4}
\end{align}
By comparing (\ref{y21gauge_2}) and (\ref{is2s4})
we find the last relation in (\ref{y21relations}) holds.

\subsection{$L^{1,2,1}$ (suspended pinch point)}\label{spp.sec}
\begin{figure}[htb]
\centering
\includegraphics{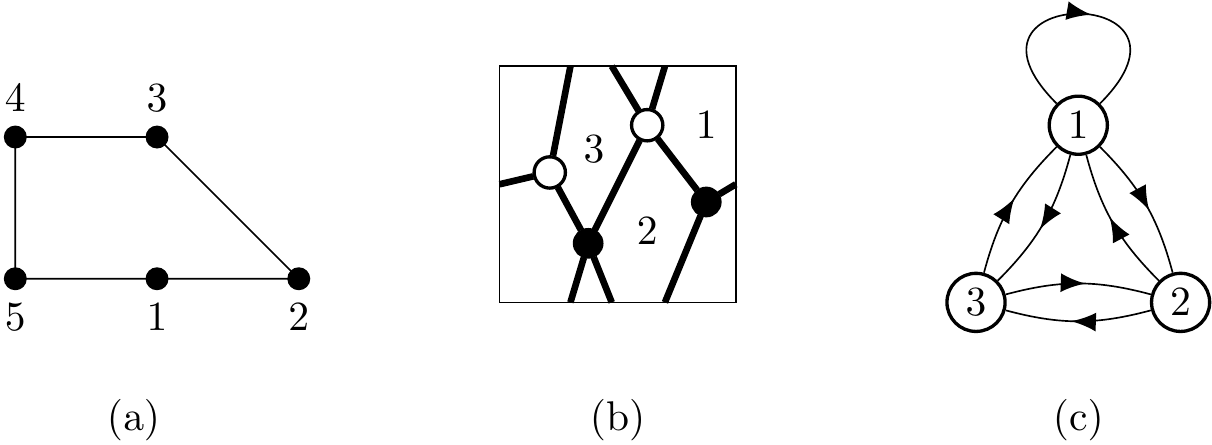}
\caption{The toric diagram of $L^{1,2,1}$ (a), the corresponding bipartite graph (b), and the corresponding quiver diagram (c) are shown.}
\label{spp}
\end{figure}

The final example is $L^{1,2,1}$,
the base of the suspended pinch point.
The toric diagram,
the bipartite graph,
and the quiver diagram
are shown in Figure \ref{spp}.
\begin{table}
\caption{Matter contents and charge assignments are shown for the $L^{1,2,1}$ model.
The charges $R_1$, $R_1'$ and $\wt B$
are related to the shrinking cycle and
are defined in Appendix \ref{refinedb2.sec}.
}
\label{sppcharges}
\vspace{-0.5cm}
\begin{align}
\begin{array}{cccccccccccc}\hline\hline
{\rm Fields} &R_1 & R'_1 &R_2	&R_3	&R_4	&R_5 	&r	& F_1 	&F_2 	&NB	&N\wt B	\\\hline
X_{11} &0 &0 &0 &1 &1 &0 &1   &0  &-2 &0  &0  \\
X_{12} &0 &1 &0 &0 &0 &1 &1/2 &0  &1 &-1 &-1 \\
X_{21} &1 &0 &1 &0 &0 &0 &1/2 &0  &1 &1 &1 \\
X_{23} &0 &0 &0 &0 &1 &0 &1/2 &-1 &-1 &2 &0  \\
X_{32} &0 &0 &0 &1 &0 &0 &1/2 &1 &-1 &-2 &0  \\
X_{31} &1 &0 &0 &0 &0 &1 &1/2 &0  &1 &-1 &1 \\
X_{13} &0 &1 &1 &0 &0 &0 &1/2 &0  &1 &1 &-1 \\\hline
\end{array}\nn
\end{align}
\end{table}
Because the toric diagram
has a vertex on the boundary which is not a corner
(vertex $1$ in Figure \ref{spp} (a)),
the manifold has the corresponding shrinking cycle.
In this section
we neglect the wrapping number on the shrinking cycle
by setting the corresponding fugacity to be $1$.
See Appendix \ref{refinedb2.sec} for an analysis with it taken into
account.
In this section
we take account of only four charges $R_2$, $R_3$, $R_4$, and $R_5$
associated with the corners of the toric diagram.
For convenience we define the following four linear combinations.
\begin{align}
r&=\frac{1}{2}(R_2+R_3+R_4+R_5),\nonumber\\
F_1&=R_3-R_4,\nonumber\\
F_2&=R_2-R_3-R_4+R_5,\nonumber\\
B&=\frac{1}{N}(R_2-2R_3+2R_4-R_5).
\label{spp_fourq}
\end{align}
$r$ is not the $R$-charge $r_*$ appearing in the superconformal algebra,
which is given by \cite{Benvenuti:2005ja,Franco:2005sm,Butti:2005sw}
\begin{align}
r_*=\frac{1}{\sqrt{3}}(R_2+R_5)+\left(1-\frac{1}{\sqrt{3}}\right)(R_3+R_4).
\label{spp_scr}
\end{align}
The charge assignments for charges defined above are shown
in Table \ref{sppcharges}.

In this example
we have $\ZZ_2$ symmetry acting on the toric diagram
as the permutation of the vertices $(1,2,3,4,5)\rightarrow(1,5,4,3,2)$.
On the gauge theory side this is the charge conjugation mapping a
bi-fundamental field $X_{ab}$ to $X_{ba}$.
$r_*$ and $F_2$ are even under this charge conjugation while
$F_1$ and $B$ are odd.

The gauge theory is non-chiral
and the baryonic symmetry is
obtained by replacing the $SU(N)$ gauge groups
by $U(1)$ and removing the diagonal $U(1)$.
There are two baryonic charges, and $B$ defined
in (\ref{spp_fourq}) is one of them.
The other (shown as $\wt B$ in Table \ref{sppcharges})
is associated with the shrinking cycle and
we neglect it in this section.
We have no discrete factor in the baryonic symmetry.

We define new fugacities by
\begin{align}
\prod_{I=1}^5v_I^{R_I}
=q^{\frac{3}{2}r}u^{F_1}v^{F_2}\zeta^{B},
\end{align}
and $v_I$ are given in terms of new variables by
\begin{align}
v_1=1,\quad
v_2=q^{\frac{3}{4}}v\zeta^{\frac{1}{N}},\quad
v_3=\frac{q^{\frac{3}{4}}u}{v\zeta^{\frac{2}{N}}},\quad
v_4=\frac{q^{\frac{3}{4}}\zeta^{\frac{2}{N}}}{uv},\quad
v_5=\frac{q^{\frac{3}{4}}v}{\zeta^{\frac{1}{N}}},
\end{align}
$v_1$ is the fugacity associated with the shrinking cycle
and is set to be $1$.

\begin{table}[htb]
\caption{
The lowest and higher order contributions to the indices of the D3-branes with
the wrapping numbers $B\in\ZZ$, and the corresponding brane configurations are shown for $L^{1,2,1}$.
}
\label{spp_lowest}
\vspace{-0.5cm}
\begin{align}
\begin{array}{c | l | l}\hline\hline
B	&{\rm lowest\ order,\ (config.)}	&{\rm higher\ order,\ (config.)}	\\\hline
-2	&q^{\frac{3}{4}N},(S_3)	&q^{\frac{3}{2}N},(2S_5)	\\
-1	&q^{\frac{3}{4}N},(S_5)	&q^{\frac{3}{2}N},(S_2+S_3)		\\
0	&1					&q^{\frac{3}{2}N},(S_3+S_4,\cdots)	\\
1	&q^{\frac{3}{4}N},(S_2)	&q^{\frac{3}{2}N},(S_4+S_5)		\\
2	&q^{\frac{3}{4}N},(S_4)	&q^{\frac{3}{2}N},(2S_2)	\\
\hline
\end{array}\nn
\end{align}
\end{table}

We define the index ${\cal I}_B^{\rm gauge}$ of the sector with a specific baryonic charge $B$ by the expansion
\begin{align}
{\cal I}^{\rm gauge}=\sum_{B \in \ZZ}{\cal I}_B^{\rm gauge},\qquad
{\cal I}_B^{\rm gauge}\propto\zeta^B.
\end{align}
Brane configurations and expected orders of their contribution
are shown in Table \ref{spp_lowest} for $-2\leq B\leq 2$.
The expected relations obtained from the order estimation
in Table \ref{spp_lowest} are
\begin{align}
{\cal I}^{\rm gauge}_0
&={\cal I}^{\rm KK}+\mathcal{O}(q^{\frac{3}{2}N}),
\nn\\
\frac{{\cal I}^{\rm gauge}_{1}}{{\cal I}^{\rm KK}}
&
={\cal I}^{\rm D3}_{S_2}+\mathcal{O}(q^{\frac{3}{2}N}),
\nn\\
\frac{{\cal I}^{\rm gauge}_{-2}}{{\cal I}^{\rm KK}}
&={\cal I}^{\rm D3}_{S_3}+\mathcal{O}(q^{\frac{3}{2}N}),
\nn\\
\frac{{\cal I}^{\rm gauge}_{2}}{{\cal I}^{\rm KK}}
&={\cal I}^{\rm D3}_{S_4}+\mathcal{O}(q^{\frac{3}{2}N}),
\nn\\
\frac{{\cal I}^{\rm gauge}_{-1}}{{\cal I}^{\rm KK}}
&
={\cal I}^{\rm D3}_{S_5}+\mathcal{O}(q^{\frac{3}{2}N}).
\label{spp_expected}
\end{align}
Let us confirm the relations in (\ref{spp_expected}) for $N=2$.
See also Appendix \ref{n3.sec} for the analysis with $N=3$.

For the $B=0$ sector on the gauge theory side we obtain
\begin{align}
{\cal I}^{\rm gauge}_0 &=
1
+\left(v^2+\frac{2}{v^2}\right)q^{\frac{3}{2}}
+\left(u v+\frac{v}{u}\right)q^{\frac{9}{4}}
+\left(4v^4+\frac{5}{v^4}+2\right)q^3
+{\cal O}(q^{\frac{15}{4}}).\label{sppgauge_0}
\end{align}
On the gravity side
by using the formula (\ref{toricKK}) with
the fugacities
\begin{align}
w_{\frac{1}{2}}=w_{1+\frac{1}{2}}=v_3v_4,\quad
w_{2+\frac{1}{2}}=v_1v_4v_5^2,\quad
w_{3+\frac{1}{2}}=v_1v_2v_5,\quad
w_{4+\frac{1}{2}}=v_1v_2^2v_3,
\end{align}
we obtain the index
\begin{align}
{\cal I}^{\rm KK}
&=
(\cdots{\rm identical\ terms\ with\ (\ref{sppgauge_0})}\cdots)
+\left(2v^4+\frac{5}{v^4}+2\right)q^3
+{\cal O}(q^{\frac{15}{4}}).
\end{align}
These two indices are consistent with the first relation in (\ref{spp_expected}).

Let us discuss $B\neq0$ sectors.
Because ${\cal I}_B$ and ${\cal I}_{-B}$ are related by the charge conjugation
we only consider $B>0$ sectors.

For the $B=1$ sector we have on the gauge theory side
\begin{align}
\frac{{\cal I}^{\rm gauge}_{1}}{{\cal I}^{\rm KK}}
&=
2\zeta\left[
v^2q^{\frac{3}{2}}
+\frac{v}{u}q^{\frac{9}{4}}
+\left(\frac{1}{u^2}-1+v^2 \chi _1^J\right)q^3
-\frac{v^3}{u}q^{\frac{15}{4}}
+{\cal O}(q^{\frac{9}{2}})
\right],\label{sppgauge_1}
\end{align}
where $\chi_n^J=\chi_n(y)$,
and on the gravity side
\begin{align}
{\cal I}^{\rm D3}_{S_2}
&=2v_2^N\Pexp\left( f(q,y,w_{1+\frac{1}{2}},w_{2+\frac{1}{2}}) \right)
\nonumber\\
&\stackrel{(N=2)}{=}
2\zeta\left[
(\cdots{\rm identical\ terms\ with\ (\ref{sppgauge_1})}\cdots)
+\left(\frac{1}{u^3v}-\frac{v^3}{u}\right)q^{\frac{15}{4}}
+{\cal O}(q^{\frac{9}{2}})
\right].
\label{id3s2spp}
\end{align}
These two are consistent with the second relation in (\ref{spp_expected}).

For the $B=2$ sector we have on the gauge theory side
\begin{align}
\frac{{\cal I}^{\rm gauge}_{2}}{{\cal I}^{\rm KK}}
&=
\zeta^2\left[\frac{1}{u^2 v^2}q^{\frac{3}{2}}
+\frac{ v}{u}q^{\frac{9}{4}}
+\left(-\frac{1}{u^2}+3v^4+\frac{\chi _1^J}{u^2v^2}\right)q^3
+{\cal O}(q^{\frac{15}{4}})
\right],\label{sppgauge_2}
\end{align}
and on the gravity side
\begin{align}
{\cal I}_{S_4}^{\rm D3}
&=v_4^N\Pexp\left( f(q,y,w_{3+\frac{1}{2}},w_{4+\frac{1}{2}}) \right)
\nonumber\\
&\stackrel{(N=2)}{=}
\zeta^2\bigg[
(\cdots{\rm identical\ terms\ with\ (\ref{sppgauge_2})}\cdots)\nn\\
&+\left(-\frac{1}{u^2}+v^4+\frac{\chi _1^J}{u^2v^2}\right)q^3
+{\cal O}(q^{\frac{15}{4}})
\bigg].
\end{align}
These two are consistent with the forth relation in (\ref{spp_expected}).

\section{Conclusions}\label{discussion.sec}
We investigated the superconformal index of ${\cal N}=1$ quiver gauge theories
realized on D3-branes in toric Calabi-Yau manifolds.
The holographic dual of such a quiver gauge theory is
type IIB string theory in $AdS_5\times SE_5$,
where $SE_5$ is a toric Sasaki-Einstein manifold.
A D3-brane wrapped on a supersymmetric three-cycle in $SE_5$,
which corresponds to a baryonic operator  in the gauge theory,
contributes to the index as a finite $N$ correction.
We proposed the formula (\ref{id3formula}) for the correction to the index
due to such a D3-brane and fluctuation modes of massless fields on the brane.
The formula (\ref{id3formula}) is a natural generalization of similar formulas
proposed in \cite{Arai:2019xmp} for S-folds and in \cite{Arai:2019wgv} for orbifolds.
Similarly to these previous cases a wrapped D3-brane has topology $\bm{S}^3/\Gamma$,
where $\Gamma$ is an abelian group.
A difference is that for a toric manifold the $\bm{S}^3$ is in general not round.
Even so the formula is still quite simple thanks to the fact
that the index depends on the background geometry through only small number of parameters.

The formula is applicable to a general toric quiver gauge theories.
Starting from the toric data of the $SE_5$ we can easily calculate
the corrections induced by D3-branes with single wrapping.
We did not take account of D3-branes with multiple wrapping.
We confirmed that the formula works correctly for several examples
($SE_5=T^{1,1},T^{2,2},Y^{2,1}$, and $L^{1,2,1}$)
by comparing the index obtained from the formula
with the result of numerical calculation using localization method.
The errors are consistent with the interpretation
that they are due to branes with multiple wrapping.

The formula consists of three factors:
the degeneracy factor $m_I$,
the classical factor $v_I^N$,
and the excitation factor $\Pexp i^{\rm D3}_{S_I}$.
The degeneracy factor $m_I$ was interpreted
as the degeneracy of states on the wrapped D3-brane due to the presence of different gauge holonomies.
Because we considered only D3-brane with single wrapping the theory on the D3-brane is
$U(1)$ gauge theory consisting only of neutral fields.
Then holonomies do not couple to any excitation modes on the brane,
and different holonomies simply give overall numerical factor $m_I$.

A part of the degeneracy factor is associated with
torsion cycles or shrinking cycles, and we can turn on fugacities coupling to them.
In the main text we neglected wrapping on such cycles
by setting the corresponding fugacity to be 1. See Appendix \ref{refinedb.sec}
for preliminary analyses in which we introduce fugacities
to see refined structure of the index in two examples.

\section*{Acknowledgments}
We would like to thank D.~Yokoyama
for valuable discussions and comments.

\appendix
\section{Localization formula}\label{localization.sec}
A toric quiver gauge theory has gauge group
$SU(N)^{n_V}$ and bi-fundamental chiral multiplets $X_{ab}$
belonging to the representation $(N,\ol N)$ of
$SU(N)_a\times SU(N)_b$.
The superconformal index is given by
\begin{align}
{\cal I}^{\rm gauge}=\prod_{a=1}^{n_V}\int d\mu^{(a)}\Pexp i^{\rm gauge},
\end{align}
where $\Pexp$ is the plethystic exponential defined by
\begin{align}
\Pexp\left(\sum_k c_kx_1^{n_{1,k}}\cdots x_p^{n_{p,k}}\right)
=\prod_k(1-x_1^{n_{1,k}}\cdots x_p^{n_{p,k}})^{-c_k},
\label{pexpdef}
\end{align}
$\int d\mu^{(a)}$ is the integral over the $SU(N)_a$ gauge fugacities $z^{(a)}_i$ ($i=1,\ldots,N$)
defined by
\begin{align}
\int d\mu^{(a)}=\frac{1}{N!}\int\prod_{i=1}^{N-1}\frac{dz^{(a)}_i}{2\pi iz^{(a)}_i}
\prod_{i>j}(z^{(a)}_i-z^{(a)}_j).
\end{align}
The gauge fugacities are constrained by
\begin{align}
\prod_{i=1}^Nz^{(a)}_i=1.
\end{align}
The single-particle index $i^{\rm gauge}$
is the sum of contributions of vector multiplets $V^a$ and bi-fundamental chiral multiplets $X_{ab}$:
\begin{align}
i^{\rm gauge}=\sum_{a=1}^{n_V}i[V^a]+\sum_{X_{ab}}i[X_{ab}],
\end{align}
where the contribution of the $SU(N)_a$ vector multiplet $V_a$ is
\begin{align}
i[V_a]=-\left(\frac{yq^{\frac{3}{2}}}{1-yq^{\frac{3}{2}}}
+\frac{y^{-1}q^{\frac{3}{2}}}{1-y^{-1}q^{\frac{3}{2}}}\right)\chi^{(a)}_{\rm adj},
\end{align}
and the contribution of the
chiral multiplet $X_{ab}$ is
\begin{align}
i[X_{ab}]
=\frac{\chi[X_{ab}]-q^3\ol\chi[X_{ab}]}
{(1-yq^{\frac{3}{2}})(1-y^{-1}q^{\frac{3}{2}})}.
\end{align}
$\chi[X_{ab}]$ is defined by
\begin{align}
\chi[X_{ab}]
=\prod_Iv_I^{R_I[X_{ab}]}
\chi^{(a)}_{\rm fund}
\ol\chi^{(b)}_{\rm fund},
\end{align}
and $\ol\chi$ is defined from this by replacing all
fugacities with their inverse.
$\chi^{(a)}_{\rm fund}$,
$\ol\chi^{(b)}_{\rm fund}$, and $\chi^{(a)}_{\rm adj}$ are the $SU(N)_a$ characters of the
fundamental, the anti-fundamental, and the adjoint representations, respectively:
\begin{align}
\chi^{(a)}_{\rm fund}=\sum_{i=1}^Nz^{(a)}_i,\quad
\ol\chi^{(a)}_{\rm fund}=\sum_{i=1}^N\frac{1}{z^{(a)}_i},\quad
\chi_{\rm adj}^{(a)}=\chi_{\rm fund}^{(a)}\ol\chi^{(a)}_{\rm fund}-1.
\end{align}

\section{Results for $N=3$}\label{n3.sec}
In this appendix we show the results for $N=3$
for models analyzed in Section \ref{examples.sec} for $N=2$.
\subsection{$T^{1,1}$ (conifold)}
Let us confirm the relations (\ref{conirelation_0}) and (\ref{conirelations}) for $N=3$.
For the relation (\ref{conirelation_0}) the indices appearing on the left and right hand sides
are given by
\begin{align}
\mathcal{I}_0^{\rm gauge}&=
1+\chi _1^u\chi _1^vq^{\frac{3}{2}}
+(2-\chi _2^u-\chi _2^v+2\chi _2^u\chi _2^v)q^3\nn\\
&+(2\chi _1^u\chi _1^v-\chi _3^u\chi _1^v-\chi _1^u\chi _3^v+3\chi _3^u\chi _3^v)q^{\frac{9}{2}}\nn \\
&+(2-2\chi _2^u+\chi _4^u-2\chi _2^v+\chi _4^v-2\chi _4^u\chi _2^v-2\chi _u^2\chi _4^v\nn\\
&+4\chi _4^u\chi _4^v+2(\chi _1^u\chi _1^v+\chi _3^u\chi _3^v)\chi _1^J)q^6+\mathcal{O}(q^{\frac{15}{2}}),\label{conin3gauge_0}\\
\mathcal{I}^{\rm KK}&=(\cdots \text{identical terms\ with\ (\ref{conin3gauge_0})}\cdots )\nn \\
&+(6 - 2 \chi ^u_2+\chi ^u_4-2\chi ^v_2+\chi ^v_4+4\chi ^u_2\chi ^v_2-2\chi ^u_4\chi ^v_2\nn\\
&-2\chi ^u_2\chi ^v_4+5\chi ^u_4\chi ^v_4)q^6+\mathcal{O}(q^{\frac{15}{2}}),
\end{align}
and these satisfy (\ref{conirelation_0}).
The first relation of (\ref{conirelations}) is confirmed for $N=2$ with the
results
\begin{align}
\frac{\mathcal{I}_1^{\rm gauge}}{\mathcal{I}^{\rm KK}}&=\zeta \Big[\chi _3^uq^{\frac{9}{4}}+(\chi _1^v-\chi _2^u\chi _1^v+(-\chi _1^u+\chi _3^u)\chi _1^J)q^{\frac{15}{4}}\nn \\
&+(\chi _1^u-\chi _3^u-2\chi _1^u\chi _2^v+2\chi _1^v\chi _1^J+(-\chi _1^u+\chi _3^u)\chi _2^J)q^{\frac{21}{4}}\nn \\
&+(\chi _3^v-\chi _4^u\chi _3^v+(\chi _1^u-\chi _3^u-\chi _1^u\chi _2^v+\chi _3^u\chi _2^v)\chi _1^J+(-\chi _1^u+\chi _3^u)\chi _3^J)q^{\frac{27}{4}}\nn \\
&+(\chi _3^u-\chi _5^u+3\chi _1^u\chi _2^v+\chi _1^v\chi _4^v-\chi _5^u\chi _2^v-\chi _7^u\chi _4^v\nn \\
&+(-2\chi _1^v-2\chi _3^v-\chi _2^u\chi _1^v-\chi _2^u\chi _3^v+\chi _4^u\chi _1^v+\chi _6^u\chi _3^v)\chi _1^J\nn \\
&+(2\chi _1^u-2\chi _3^u+\chi _1^u\chi _2^v)\chi _2^J+(-\chi _1^u+\chi _3^v)\chi _4^J) q^{\frac{33}{4}}\nn \\
&+(\chi _1^v+3\chi _3^v-2\chi _5^v+14\chi _2^u\chi _1^v+4\chi _2^u\chi _3^v+2\chi _2^u\chi _5^v+6\chi _4^u\chi _1^v\nn \\
&+7\chi _4^u\chi _3^v+2\chi _6^u\chi _1^v+\chi _6^u\chi _3^v+\chi _6^u\chi _5^v-\chi _8^u\chi _1^v+\chi _8^u\chi _3^v-\chi _8^u\chi _5^v\nn \\
&+(-6\chi _1^u-4\chi _1^u\chi _2^v-\chi _1^u\chi _4^v-8\chi _3^u-10\chi _3^u\chi _2^v-3\chi _3^u\chi _4^v\nn \\
&+-\chi _5^u-7\chi _5^u\chi _2^v-\chi _5^u\chi _4^v+\chi _7^u-\chi _7^u\chi _4^v)\chi _1^J\nn \\
&+(-\chi _1^v-3\chi _3^v+4\chi _2^u\chi _1^v+4\chi _2^u\chi _3^v+4\chi _4^u\chi _1^v-\chi _6^u\chi _1^v+3\chi _6^u\chi _3^v)\chi _2^J\nn \\
&+(\chi _1^u-\chi _3^u)\chi _3^J+(-\chi _1^u+\chi _3^u)\chi _5^J)q^{\frac{39}{4}}+\mathcal{O}(q^{\frac{41}{4}})\Big],\label{conin3gauge_1}
\end{align}
\begin{align}
\mathcal{I}_{S_1}^{\rm D3}+\mathcal{I}_{S_3}^{\rm D3}&=\zeta \Big[(\cdots \text{identical terms\ with\ (\ref{conin3gauge_1})}\cdots )\nn \\
&+(-2\chi _1^v-\chi _5^v+2\chi _2^u\chi _1^v+2\chi _2^u\chi _3^v+\chi _2^u\chi _5^v+\chi _4^v\chi _1^v\nn \\
&+2\chi _4^u\chi _3^v+\chi _4^u\chi _5^v-\chi _6^u\chi _3^v-\chi _8^u\chi _1^v-\chi _8^u\chi _3^v-\chi _{10}^u\chi _5^v\nn \\
&+(2\chi _1^u\chi _2^v-2\chi _3^u-3\chi _3^u\chi _2^v-2\chi _3^u\chi _4^v+\chi _7^u+\chi _7^u\chi _2^v+\chi _9^u\chi _4^v)\chi _1^J\nn \\
&+(\chi _1^u-\chi _3^u)\chi _3^J+(-\chi _1^u+\chi _3^u)\chi _5^J)q^{\frac{39}{4}}+ \mathcal{O}(q^{\frac{41}{4}})\Big].
\end{align}
The second relation in (\ref{conirelations}) follows from
the first one by exchanging the flavor fugacities as in the $N=2$ case.

\subsection{$T^{2,2}$ (complex cone over $F_0$)}
Let us check the relations (\ref{f0relations}) for the $N=3$ case.
For the first relation the  explicit calculation shows
\begin{align}
\mathcal{I}_{0}^{\mathrm{gauge}}&=1+\left(\chi ^u_2 \chi ^v_2-\chi ^u_2-\chi ^v_2+1\right)q^3
+4 \left(\chi ^u_1 \chi ^v_1+\chi ^u_3 \chi ^v_3\right)q^{\frac{9}{2}}+\mathcal{O}(q^6),\label{f0n3gauge_0}\\
\mathcal{I}^{\mathrm{KK}}&=(\cdots{\rm identical\ terms\ with\ (\ref{f0n3gauge_0})}\cdots)+0q^{\frac{9}{2}}+\mathcal{O}(q^{6}),
\end{align}
and we can check the first relation in (\ref{f0relations}) holds.
To confirm the  the last relation we have
\begin{align}
\frac{\mathcal{I}_{1}^{\mathrm{gauge}}}{\mathcal{I}^{\mathrm{KK}}}&=\zeta\Big[2 \chi ^u_3q^{\frac{9}{4}}+2\left(\chi ^u_3-\chi ^u_1\right) \chi ^J_1q^{\frac{15}{4}}
+2\left( -\chi ^u_3- \chi ^u_1\chi ^v_2+\left(\chi ^u_3 -\chi^u_1\right)\chi ^J_2\right)q^{\frac{21}{4}} \nn\\
&+2 \left(3 \chi ^u_4 \chi ^v_1+3 \chi ^u_6 \chi ^v_3+3 \chi ^u_2 \left(\chi ^v_1+\chi ^v_3\right)\right.\nn\\
&\left.+\left(\chi ^u_1 \chi ^v_2 +\chi ^u_1 -\chi ^u_3 \right)\chi
   ^J_1-\left(\chi ^u_1 -\chi ^u_3 \right)\chi ^J_3\right)q^{\frac{27}{4}}+\mathcal{O}(q^{\frac{29}{4}})\Big],\label{f0n3gauge_1}\\
\mathcal{I}_{S_1}^{D3}+\mathcal{I}_{S_3}^{D3}&=\zeta\Big[(\cdots{\rm identical\ terms\ with\ (\ref{f0n3gauge_1})}\cdots)\nn\\
&+2 \left(\left(\chi ^u_1 \chi ^v_2 +\chi ^u_1 -\chi ^u_3 \right)\chi ^J_1-\left(\chi ^u_1 -\chi ^u_3 \right)\chi ^J_3\right)q^{\frac{27}{4}} +\mathcal{O}(q^{\frac{33}{4}})\Big].
\end{align}
Again the second relation in (\ref{f0relations}) can be checked by fugacities transformation (\ref{conifugarelation}).

\subsection{$Y^{2,1}$ (complex cone over $dP_1$)}
We check the relations (\ref{y21relations}) for the $N=3$ case except for the last one.
For the last case the calculation on the gauge theory side takes  much time for an ordinary laptop computer and we could not finish the computation.
The first  relation in (\ref{y21relations}) can be confirmed by
\begin{align}
\mathcal{I}_{0}^{\mathrm{gauge}}&=1+ v \chi ^u_1q^{\frac{9}{4}}+\left(\frac{\chi ^u_3}{v}-\frac{\chi ^u_1}{v}\right)q^{\frac{15}{4}}+  \left(2 v^2\chi
   ^u_2-v^2\right)q^{\frac{9}{2}}+ \left(\chi ^u_4-1\right)q^6\nn\\
   &- \left(v^3\chi ^u_1-11v^3 \chi^u_3\right)q^{\frac{27}{4}}+\mathcal{O}(q^{\frac{29}{4}}),\label{y21n3gauge_0}\\
\mathcal{I}^{\mathrm{KK}}&=(\cdots{\rm identical\ terms\ with\ (\ref{y21n3gauge_0})}\cdots)+ \left(3 v^3 \chi ^u_3-v^3 \chi ^u_1\right)q^{\frac{27}{4}}+\mathcal{O}\left(q^{{\frac{15}{2}}}\right).
\end{align}

For the second relation in (\ref{y21relations}) we can check this actually holds.
\begin{align}
\frac{\mathcal{I}_{-1}^{\mathrm{gauge}}}{\mathcal{I}^{\mathrm{KK}}}&=\frac{1}{\zeta}\Big[3 v^{\frac{3}{2}}q^{\frac{9}{4}}+3 v^{\frac{1}{2}} \chi ^u_1q^3
+3\left(\frac{\chi ^u_2}{v^{\frac{1}{2}}}+v^{\frac{3}{2}} \chi
   ^J_1\right) q^{\frac{15}{4}}+\frac{3 \chi ^u_3}{v^{\frac{3}{2}}}q^{\frac{9}{2}} \nn\\
   &+3\left( - v^{\frac{3}{2}}-\frac{ \chi ^J_1}{v^{\frac{1}{2}}}+v^{\frac{3}{2}} \chi
   ^J_2\right)q^{\frac{21}{4}}+\mathcal{O}(q^6)\Big],\label{y21n3gauge_-1}\\
\mathcal{I}_{S_1}^{D3}&=\frac{1}{\zeta}\Big[(\cdots{\rm identical\ terms\ with\ (\ref{y21n3gauge_-1})}\cdots)\nn\\
&+3 \left(\frac{\chi ^u_4}{v^{\frac{5}{2}}}-v^{\frac{3}{2}}-\frac{\chi ^J_1}{v^{\frac{1}{2}}}+v^{\frac{3}{2}} \chi ^J_2\right)q^{\frac{21}{4}} +\mathcal{O}(q^6)\Big].
\end{align}
For the third relation in (\ref{y21relations}) the explicit calculation shows
\begin{align}
\frac{\mathcal{I}_{-3}^{\mathrm{gauge}}}{\mathcal{I}^{\mathrm{KK}}}
&=\frac{1}{\zeta^3}\bigg[\frac{q^{\frac{9}{4}}}{v^{\frac{3}{2}}}+\left(v^{\frac{1}{2}}+\frac{\chi ^J_1}{v^{\frac{3}{2}}}\right)q^{\frac{15}{4}}
-\frac{\chi ^u_1}{v^{\frac{1}{2}}}q^{\frac{9}{2}}
+\left(v^{\frac{5}{2}}+\frac{\chi ^J_2}{v^{\frac{3}{2}}}-\frac{1}{v^{\frac{3}{2}}}\right)q^{\frac{21}{4}}\nn\\
&+\left(-v^{\frac{1}{2}} \chi^u_2+10 v^{\frac{9}{2}}-\frac{\chi ^J_1}{v^{\frac{3}{2}}}+\frac{\chi ^J_3}{v^{\frac{3}{2}}}\right)q^{\frac{27}{4}}
+\mathcal{O}(q^{\frac{29}{4}})\bigg],\label{y21n3gauge_-3}\\
\mathcal{I}_{S_1}^{D3}&=\frac{1}{\zeta^3}\Big[(\cdots{\rm identical\ terms\ with\ (\ref{y21n3gauge_-3})}\cdots)\nn\\
&+ \left(-v^{\frac{1}{2}} \chi ^u_2+v^{\frac{9}{2}}-\frac{\chi ^J_1}{v^{\frac{3}{2}}}+\frac{\chi ^J_3}{v^{\frac{3}{2}}}\right)q^{\frac{27}{4}}+\mathcal{O}(q^{\frac{15}{2}})\Big].
\end{align}

\subsection{$L^{1,2,1}$ (suspended pinch point)}
The first relation in (\ref{spp_expected}) is confirmed
for $N=3$ by comparing the following
\begin{align}
{\cal I}^{\rm gauge}_0
&=
1
+\left(v^2+\frac{2}{v^2}\right)q^{\frac{3}{2}}
+ \left(u v+\frac{v}{u}\right)q^{\frac{9}{4}}
+\left(2 v^4+\frac{5}{v^4}+2\right)q^3
\nn\\&\quad
+ \left(u v^3+\frac{2u}{v}+\frac{v^3}{u}+\frac{2}{uv}\right)q^{\frac{15}{4}}
\nn\\&\quad
+ \left(2 u^2 v^2+\frac{2 v^2}{u^2}+5v^6+\frac{10}{v^6}+5 v^2+\frac{5}{v^2}\right)q^{\frac{9}{2}}
+{\cal O}(q^{\frac{21}{4}}),\label{sppn3gauge_0}
\\
{\cal I}^{\rm KK}
&=
(\cdots{\rm identical\ terms\ with\ (\ref{sppn3gauge_0})}\cdots)
\nn\\&\quad
+ \left(2 u^2 v^2+\frac{2 v^2}{u^2}+3v^6+\frac{10}{v^6}+5 v^2+\frac{5}{v^2}\right)q^{\frac{9}{2}}
+{\cal O}(q^{\frac{21}{4}}).
\end{align}

The second relation in (\ref{spp_expected}) is confirmed
by comparing the following
\begin{align}
\frac{{\cal I}^{\rm gauge}_{1}}{{\cal I}^{\rm KK}}
&=
2\zeta\left[
 v^3q^{\frac{9}{4}}
+\frac{v^2}{u}q^3
+\left(\frac{v}{u^2}-v+v^3 \chi _1^J\right)q^{\frac{15}{4}}
+\left(\frac{1}{u^3}-\frac{v^4}{u}\right)q^{\frac{9}{2}}\right.
\nn\\&\left.
+ \left(-v^3-\frac{1}{v}+v^3 \chi _2^J\right)q^{\frac{21}{4}}
+{\cal O}(q^6)\right],\label{sppn3gauge_1}
\\
{\cal I}^{\rm D3}_{S_2}
&=
2\zeta\Big[
(\cdots{\rm identical\ terms\ with\ (\ref{sppn3gauge_1})}\cdots)\nn\\
&+\left(\frac{1}{u^4v}-v^3-\frac{1}{v}+v^3 \chi _2^J\right)q^{\frac{21}{4}}
+{\cal O}(q^6)\Big].
\end{align}

The forth relation in (\ref{spp_expected}) is confirmed
by comparing the following
\begin{align}
\frac{{\cal I}^{\rm gauge}_{2}}{{\cal I}^{\rm KK}}
&=
\zeta^2\Big[
\frac{1}{u^3 v^3}q^{\frac{9}{4}}
+\frac{1}{u^2}q^3
+\left(-\frac{1}{u^3v}+\frac{v^3}{u}+\frac{\chi _1^J}{u^3 v^3}\right)q^{\frac{15}{4}}
\nn\\&
+ \left(3v^6-\frac{1}{u^2 v^2}\right)q^{\frac{9}{2}}
+{\cal O}(q^{\frac{21}{4}} )\Big],\label{sppn3gauge_2}
\\
{\cal I}^{\rm D3}_{S_4}
&=
\zeta^2\Big[
(\cdots{\rm identical\ terms\ with\ (\ref{sppn3gauge_2})}\cdots)\nn\\
&
+ \left(v^6-\frac{1}{u^2 v^2}\right)q^{\frac{9}{2}}
+{\cal O}(q^{\frac{21}{4}} )\Big].
\end{align}

\section{Refined baryonic charges}\label{refinedb.sec}
We have seen in the main text that we have two types of cycles:
visible cycles and vanishing cycles.
Correspondingly, we can divide $U(1)$ baryonic symmetries
into two classes.
We also found in an example ($T^{1,1}$) that the discrete factor
may appear in the baryonic symmetry.
In the main text we considered only the $U(1)$
symmetries associated with visible cycles,
and neglected the others by setting
the corresponding fugacities to be $1$.
In this appendix we give some analysis with
the full baryonic symmetry taken into account.

\subsection{Discrete baryonic symmetry in the $T^{2,2}$ model}\label{refinedb1.sec}

In Section \ref{f0.sec} we mentioned that the baryonic symmetry of the $T^{2,2}$ model has
the non-trivial discrete factor $G_{\rm disc}=\ZZ_2\times\ZZ_2$.
In this appendix we discuss
how the corresponding charges appear on the gravity side.

Let us first explicitly derive the baryonic symmetry
including the discrete factor.
On the gauge theory side,
the baryonic symmetry is obtained by replacing the four $SU(N)$ gauge groups
by $U(1)$.
Because the theory is chiral the $U(1)^4$ symmetry is broken to its subgroup
by anomaly.
We denote an element of the $U(1)^4$ as $(\zeta_1,\zeta_2,\zeta_3,\zeta_4)$.
Let $\beta_i$ ($i=1,2,3,4$) be the $U(1)_i$ charges, which are classically conserved.
We use the labelling shown in Figure \ref{F0toric} (c).
The conservation law is broken by the instanton effect as
\begin{align}
\Delta \beta_i=\frac{1}{N}\times 2N\times n_{i+1}
-\frac{1}{N}\times 2N\times n_{i-1}
=2n_{i+1}-2n_{i-1},
\end{align}
where $n_i$ are the $SU(N)_i$ instanton numbers.
For an element $(\zeta_1,\zeta_2,\zeta_3,\zeta_4)$ to be
anomaly-free, the following relation must hold for arbitrary $n_i$:
\begin{align}
1=\zeta_1^{\Delta \beta_1}
\zeta_2^{\Delta \beta_2}
\zeta_3^{\Delta \beta_3}
\zeta_4^{\Delta \beta_4}
=\left(\frac{\zeta_4}{\zeta_2}\right)^{2n_1}
\left(\frac{\zeta_1}{\zeta_3}\right)^{2n_2}
\left(\frac{\zeta_2}{\zeta_4}\right)^{2n_3}
\left(\frac{\zeta_3}{\zeta_1}\right)^{2n_4}.
\end{align}
Namely, $\zeta_i$ must satisfy
\begin{align}
\frac{\zeta_2^2}{\zeta_4^2}=\frac{\zeta_3^2}{\zeta_1^2}=1.
\label{anomfree}
\end{align}
We can set $\zeta_4=1$ with the decoupled diagonal $U(1)$.
Then the solution of (\ref{anomfree}) is
\begin{align}
\zeta_1=\sigma_1\zeta, \quad \zeta_2=\sigma_2, \quad \zeta_3=\zeta, \quad \zeta_4=1,
\label{fourzeta}
\end{align}
where $\zeta\in U(1)$ and $\sigma_1,\sigma_2=\pm1$.
This means that the anomaly free baryonic symmetry is $U(1)\times \ZZ_2^2$.
$\zeta$ is identified with the fugacity defined in (\ref{f0fug}).
In addition, we can introduce $\ZZ_2$ valued fugacities $\sigma_1$ and $\sigma_2$
to define the index.

The action of each $\ZZ_2$ symmetry on bi-fundamental fields
is read off from (\ref{fourzeta}).
They act non-trivially only on baryonic operators.
Corresponding to the four arrows in the quiver diagram in Figure \ref{F0toric} (c)
we have four elementary baryonic operators with dimension $\frac{3}{4}N$:
\begin{align}
B_{12}=\det X_{12},\quad
B_{23}=\det X_{23},\quad
B_{34}=\det X_{34},\quad
B_{41}=\det X_{41},
\end{align}
and these carry $\ZZ_2$ charges shown in Table \ref{F0_z2charges}.
(We neglected the superscripts of bi-fundamental fields,
which play no role here.)
\begin{table}
\caption{Baryonic charges of the elementary baryonic operators are shown for the $F_0$ model.
$B$ denotes the $U(1)$ charge and $b_1$ and $b_2$ denote the $\ZZ_2$ charges. }
\label{F0_z2charges}
\begin{center}
\begin{tabular}{cccc}
\hline
\hline
operators & $B$ & $b_1$  & $b_2$ \\
\hline
$B_{12}$ & $+1$ & $1$ & $1$ \\
$B_{23}$ & $-1$ & $0$ & $1$ \\
$B_{34}$ & $+1$ & $0$ & $0$ \\
$B_{41}$ & $-1$ & $1$ & $0$ \\
\hline
\end{tabular}
\end{center}
\end{table}

We expand the index
by using two $\ZZ_2$-valued baryonic charges together with the integer baryonic charge:
\begin{align}
\label{if0sigma}
{\cal I}^{\rm gauge}=\sum_{B\in \mathbb{Z}}\sum_{b_1,b_2= 0,1} \sigma_1^{b_1} \sigma_2^{b_2} {\cal I}_{(B;b_1,b_2)}^{\rm gauge}.
\end{align}
Unlike $\zeta$ we do not include $\sigma_1$ and $\sigma_2$ in the definition of ${\cal I}_{(B;i,j)}^{\rm gauge}$.

As we mentioned in Section \ref{f0.sec} the theory has the $\ZZ_4$ symmetry
rotating the toric diagram and the bipartite graph.
This acts on the quiver diagram as $\ZZ_4$ rotation,
and we can read off its action on the
baryonic charges from (\ref{fourzeta}) as
\begin{align}
B\rightarrow B'=-B,\quad
b_1\rightarrow b_1'=b_2+B,\quad
b_2\rightarrow b_2'=b_1.
\label{z4sym}
\end{align}
The four elementary baryonic operators
are transformed as $B_{a-1,a}\rightarrow B_{a,a+1}$.

Let us first consider the sector with charges $(B;b_1,b_2)=(0;0,0)$.
The dimension of a product of baryonic operators
contributing to this sector is at least $3N$,
and the expected relation between indices is
\begin{align}
&{\cal I}^{\rm gauge}_{(0;0,0)}
={\cal I}^{\rm KK}+\mathcal{O}(q^{3N}).
\label{f0mesexpected1}
\end{align}
Thanks to the refinement, the order of the expected error
is larger than the previous one in (\ref{f0relations}).
Indeed, the results of numerical calculation for $N=2$ is
\begin{align}
\label{f0gaugez2_0}
\mathcal{I}_{(0;0,0)}^{\mathrm{gauge}}&=1+\left(\chi ^u_2\chi ^v_2-\chi ^u_2-\chi ^v_2+1\right)q^3 \nn\\
&+\left(2\chi ^u_2  \chi ^v_2-2 \chi ^u_2 \chi ^v_4-\chi ^u_2 -2\chi ^u_4 \chi ^v_2+6\chi ^u_4 \chi ^v_4+5\chi ^u_4-\chi
   ^v_2+5 \chi ^v_4+6\right)q^6
\nonumber\\&
+\mathcal{O}(q^{\frac{29}{4}}),
\end{align}
and the corresponding Kaluza-Klein mode contribution is
\begin{align}
{\cal I}^{\rm KK}&=(\cdots \text{identical terms with (\ref{f0gaugez2_0})}\cdots)\nn\\
&+\left(2 \chi ^u_2 \chi ^v_2-2\chi ^u_2  \chi ^v_4-\chi ^u_2 -2\chi ^u_4 \chi ^v_2+2\chi ^u_4 \chi ^v_4+\chi ^u_4-\chi ^v_2+\chi
   ^v_4+2\right)q^6
\nonumber\\&
+\mathcal{O}(q^9).
\end{align}
From these relations we find that the relation (\ref{f0mesexpected1}) holds.

Next, we consider baryonic sectors with $B=\pm1$.
Let us focus on the $B=1$ sectors.
Two baryonic operators $B_{12}$ and $B_{34}$ carries charges
$(+1;1,1)$ and $(+1;0,0)$.
The corresponding brane configurations are D3-branes wrapped on the
three-cycles $S_1$ or $S_3$.
Note that D3-branes wrapped on $S_1$ and $S_3$
can be continuously deformed to each other,
and we cannot relate two cycles with two baryonic sectors one by one.
Instead, we should interpret the existence of the two sectors
as the existence of the two different values of
the $U(1)$ holonomy on a wrapped D3-brane
with the topology $\bm{S}^3/\ZZ_2$.
Because the holonomy does not couple with
any fluctuation modes on the single D3-brane
two sectors must give the same index.
This is consistent with the relation
\begin{align}
\mathcal{I}_{(1;0,0)}^{\mathrm{gauge}}
=\mathcal{I}_{(1;1,1)}^{\mathrm{gauge}}.
\end{align}
which follows from the $\ZZ_4$ symmetry.

The lowest order contribution to the sector with
$(B;\sigma_1,\sigma_2)=(1;0,0)$
comes from $B_{34}$,
and on the gravity side this corresponds to
a brane wrapping on the $S_1$ or $S_3$.
At the order of $q^{\frac{15}{4}N}$, operators like $B_{34}B_{23}^2$
contribute to the index,
and this corresponds to multiple-wrapping configurations.
Therefore the expected relation is
\begin{align}
&\frac{{\cal I}^{\rm gauge}_{(1;0,0)}}{{\cal I}^{\rm KK}}
=\frac{1}{2}\left({\cal I}^{\rm D3}_{S_1}+{\cal I}^{\rm D3}_{S_3}\right)+\mathcal{O}(q^{\frac{15}{4}N}),
\label{f0refine}
\end{align}
where the factor $1/2$ is inserted to
remove the degeneracy factor $m_I=2$
for two possible values of the holonomy
because the sector corresponds to
one of them.

The calculation on the gauge theory side with $N=2$ gives
\begin{align}\label{f0gaugez2_1}
\frac{\mathcal{I}_{(1;0,0)}^{\mathrm{gauge}}}{\mathcal{I}^{\mathrm{KK}}}
&= \zeta\Big[\chi ^u_2q^{\frac{3}{2}}
+\left(\chi ^u_2-1\right) \chi ^J_1q^3
+\left(-\chi ^u_2 -2 \chi
   ^v_2-1+\left(\chi ^u_2-1 \right)\chi ^J_2\right)q^{\frac{9}{2}}\nn\\
&+ \left(\left(\chi^u_2\chi^v_2-\chi^u_2-\chi^v_2+1\right) \chi ^J_1+ \left(\chi ^u_2-1\right)\chi ^J_3\right)q^6\nn\\
   &+\left(
   5 \chi ^u_6 \chi ^v_4-\chi ^u_4\chi ^v_2-\chi ^u_4\chi ^v_4-\chi ^u_4
   -4\chi ^u_2 \chi ^v_2+5 \chi ^u_2\chi ^v_4+5\chi ^u_2+5 \chi ^u_6\right.\nn\\
   &\left. +7 \chi ^v_2+\chi ^v_4+6+\left(-\chi ^u_2+1\right)\chi^J_2+\left(\chi ^u_2-1\right)\chi ^J_4
  \right)q^{\frac{15}{2}} +\mathcal{O}(q^{\frac{33}{4}})\Big].
\end{align}
On the gravity side we have
\begin{align}
\frac{1}{2}\left({\cal I}^{\rm D3}_{S_1}+{\cal I}^{\rm D3}_{S_3}\right)
&=\zeta\Big[(\cdots\text{identical terms with (\ref{f0gaugez2_1})}\cdots)\nn\\
&+ \left(
-\chi ^u_4 \chi ^v_2-\chi ^u_4 \chi ^v_4+\chi ^u_2
   \chi ^v_2-\chi ^u_4+2 \chi ^v_2+\chi ^v_4+1\right.\nn\\
   &\left.+ \left(-\chi ^u_2 +1 \right)\chi ^J_2 +\left(\chi ^u_2 -1\right)\chi ^J_4
\right)q^{\frac{15}{2}}+\mathcal{O}(q^9)\Big],
\end{align}
and we find that (\ref{f0refine}) holds.

Before ending this subsection
we comment on the fact that there are two $\ZZ_2$ factors
in the homology of $T^{2,2}$:
\begin{align}
H_0=\ZZ,\quad
H_1=\ZZ_2,\quad
H_2=\ZZ,\quad
H_3=\ZZ\times\ZZ_2,\quad
H_4=0,\quad
H_5=\ZZ.
\end{align}
It seems natural to identify the torsion
subgroup of the homology with
the discrete part of the
baryonic symmetry.
We leave detailed analysis of this problem
for future work.

\subsection{Shrinking cycle in the $L^{1,2,1}$ model}\label{refinedb2.sec}

As is mentioned in Section \ref{spp.sec} we have two perfect matchings
for the vertex $1$  in the toric diagram of $L^{1,2,1}$
(Figure \ref{spp} (a))
and correspondingly we have two $R$-charges $R_1$ and $R_1'$
associated with this vertex.
The charge assignments of these charges are shown
in Table \ref{sppcharges}.
They are not independent but related by
\begin{align}
R_1'=R_5-R_1+R_2,
\label{r1r1p}
\end{align}
and we have ambiguity in the definition of $R_I$.
In the main text we neglect this ambiguity by setting $v_1=1$.
In this appendix we perform an analysis with this taken into account.

\begin{figure}[htb]
\centering
\includegraphics{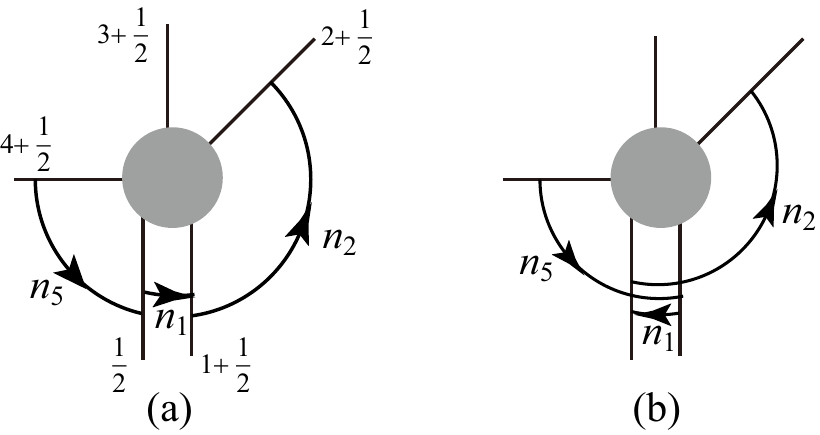}
\caption{The web-diagram of the $L^{1,2,1}$ model:
(a) wrapping numbers defined with a resolution of overlapping legs,
(b) wrapping numbers after two downgoing legs are swapped.}\label{swap.eps}
\end{figure}

For the following arguments the web diagrams are useful.
A web diagram consists of semi-infinite lines perpendicular to edges of the toric diagram.
(Figure \ref{swap.eps}).
(In the arguments below only external lines are relevant
and the central part is expressed as a blob in the figure.)
We label external lines by $r=1/2,1+1/2,\ldots$ just like edges.
These lines expresses NS5-branes in the
five-brane realization of the gauge theory.
A D3-brane wrapped on a three cycle $S_I$ is mapped to
an open D-string from the external line $r=I-1/2$ to the external line $r+1$
\cite{Imamura:2006ub,Imamura:2006ie}.
If there are more than one edges between two adjacent corners
$I_1$ and $I_2=I_1+k$, the corresponding $k$ external lines are parallel to each other.
In the conformal limit in which all the external lines meet at a point
these parallel lines coincide, and the corresponding cycles shrink.
To make these cycles visible we need to resolve the singularity
by separating parallel lines from each other.
The permutation group $S_k$ acting on these lines
is identified with the Weyl group associated with the $A_{k-1}$ singularity
acting on the shrinking cycles at the singularity.

In the case of the $L^{1,2,1}$ model
we have two parallel external lines
corresponding to the $A_1$ singularity.
We focus on the shrinking cycle $S_1$ and the visible cycles $S_5$ and $S_2$ intersecting with $S_1$.
Let $n_5$, $n_1$, and $n_2$ be the wrapping numbers for these cycles.
In the web-diagram these numbers are interpreted as the numbers of open strings (Figure \ref{swap.eps} (a)).
Let us see what happens when we swap the
two parallel external lines $1/2$ and $1+1/2$.
This corresponds to the Weyl reflection of the $A_1$.
The wrapping numbers $n_5$ and $n_2$ are kept unchanged
while $n_1$ is non-trivially transformed (Figure \ref{swap.eps} (b)):
\begin{align}
n_1\rightarrow n_1'=n_5-n_1+n_2.
\end{align}
We can easily check this generates $\ZZ_2$.
This relation is identical to the relation of $R$-charges (\ref{r1r1p}).

The four charges $r$, $F_1$, $F_2$, and $B$ defined in (\ref{spp_fourq})
do not include either $R_1$ or $R_1'$
and are invariant under the Weyl reflection.
As the additional charge associated with the shrinking cycle
it is convenient to use the Cartan generator of the $A_1$ algebra,
which is odd under the Weyl reflection. It is
\begin{align}
\wt B =\frac{1}{N}(R_1-R_1').
\end{align}
We define the corresponding fugacity $\wt\zeta$ by
\begin{align}
\prod_{I=1}^5v_I^{R_I}
=q^{\frac{3}{2}r}u^{F_1}v^{F_2}\zeta^{B}\wt\zeta^{\wt B}.
\end{align}
The fugacities $v_I$ are given by
\begin{align}
v_1=\wt\zeta^{\frac{2}{N}},\quad
v_2=\frac{q^{\frac{3}{4}}v\zeta^{\frac{1}{N}}}{\wt\zeta^{\frac{1}{N}}},\quad
v_3=\frac{q^{\frac{3}{4}}u}{v\zeta^{\frac{2}{N}}},\quad
v_4=\frac{q^{\frac{3}{4}}\zeta^{\frac{2}{N}}}{uv},\quad
v_5=\frac{q^{\frac{3}{4}}v}{\zeta^{\frac{1}{N}}\wt\zeta^{\frac{1}{N}}}.
\end{align}

A D3-brane wrapped on the shrinking cycle $S_1$
carries $\wt B=2$ and its Weyl reflection carries $\wt B=-2$.
These corresponds to the root vectors of $A_1$.
A D3-brane wrapped on the visible cycle $S_2$ carries $\wt B=-1$
and its Weyl reflection $S_2+S_1$ carries $\wt B=1$.
Namely, these two form the fundamental representation of $A_1$.
Both these two cycles contribute to ${\cal I}_{S_2}^{\rm D3}$ in (\ref{id3s2spp}),
and the refinement by introducing $\wt\zeta$ splits it into two contributions
${\cal I}'^{\rm D3}_{S_2}\propto v_2^N\propto\wt\zeta^{-1}$ and
${\cal I}'^{\rm D3}_{S_2+S_1}\propto v_2^Nv_1^N\propto\wt\zeta$.
Therefore, the degeneracy factor $2$ in (\ref{id3s2spp})
is replaced by the $A_1$ character $\chi_1(\wt\zeta)=\wt\zeta+1/\wt\zeta$:
\begin{align}
{\cal I}^{\rm D3}_{S_2}\rightarrow
{\cal I}'^{\rm D3}_{S_2}
+{\cal I}'^{\rm D3}_{S_2+S_1}
&=\frac{1}{2}\chi_1(\wt\zeta)
{\cal I}^{\rm D3}_{S_2}.
\end{align}
This is also the case for the other adjacent visible cycle $S_5$ and
its Weyl reflection $S_5+S_1$.
\begin{align}
{\cal I}^{\rm D3}_{S_5}\rightarrow
{\cal I}'^{\rm D3}_{S_5}
+{\cal I}'^{\rm D3}_{S_5+S_1}
&=\frac{1}{2}\chi_1(\wt\zeta)
{\cal I}^{\rm D3}_{S_5}.
\end{align}
By the numerical calculation on the gauge theory side
we can find the same refinement.
Namely, the overall factor $2$ on the gauge theory side
is also replaced by the character $\wt\zeta_1$
and the second and the fifth relations in (\ref{spp_expected})
still hold after the refinement.

In general,
if there are $k$ parallel external lines, the factor $k$ in the degeneracy factor
is replaced by the fundamental $A_{k-1}$ character for one of the adjacent cycle
and by the anti-fundamental $A_{k-1}$ character for the other adjacent cycle.



\begin{thebibliography}{99}
\bibitem{Maldacena:1997re}
  J.~M.~Maldacena,
  ``The Large N limit of superconformal field theories and supergravity,''
  Int.\ J.\ Theor.\ Phys.\  {\bf 38}, 1113 (1999)
  [Adv.\ Theor.\ Math.\ Phys.\  {\bf 2}, 231 (1998)]
  doi:10.1023/A:1026654312961, 10.4310/ATMP.1998.v2.n2.a1
  [hep-th/9711200].
\bibitem{Witten:1998qj}
  E.~Witten,
  ``Anti-de Sitter space and holography,''
  Adv.\ Theor.\ Math.\ Phys.\  {\bf 2}, 253 (1998)
  doi:10.4310/ATMP.1998.v2.n2.a2
  [hep-th/9802150].
\bibitem{Gubser:1998bc}
  S.~S.~Gubser, I.~R.~Klebanov and A.~M.~Polyakov,
  ``Gauge theory correlators from noncritical string theory,''
  Phys.\ Lett.\ B {\bf 428}, 105 (1998)
  doi:10.1016/S0370-2693(98)00377-3
  [hep-th/9802109].
\bibitem{Kinney:2005ej}
  J.~Kinney, J.~M.~Maldacena, S.~Minwalla and S.~Raju,
  ``An Index for 4 dimensional super conformal theories,''
  Commun.\ Math.\ Phys.\  {\bf 275}, 209 (2007)
  doi:10.1007/s00220-007-0258-7
  [hep-th/0510251].
\bibitem{Witten:1998xy}
  E.~Witten,
  ``Baryons and branes in anti-de Sitter space,''
  JHEP {\bf 9807}, 006 (1998)
  doi:10.1088/1126-6708/1998/07/006
  [hep-th/9805112].
\bibitem{Arai:2019xmp}
  R.~Arai and Y.~Imamura,
  ``Finite $N$ Corrections to the Superconformal Index of S-fold Theories,''
  arXiv:1904.09776 [hep-th].
\bibitem{Arai:2019wgv}
  R.~Arai, S.~Fujiwara, Y.~Imamura and T.~Mori,
  ``Finite $N$ corrections to the superconformal index of orbifold quiver gauge theories,''
  arXiv:1907.05660 [hep-th].
\bibitem{Feng:2000mi}
  B.~Feng, A.~Hanany and Y.~H.~He,
  ``D-brane gauge theories from toric singularities and toric duality,''
  Nucl.\ Phys.\ B {\bf 595}, 165 (2001)
  doi:10.1016/S0550-3213(00)00699-4
  [hep-th/0003085].
\bibitem{Feng:2002zw}
  B.~Feng, S.~Franco, A.~Hanany and Y.~H.~He,
  ``Symmetries of toric duality,''
  JHEP {\bf 0212}, 076 (2002)
  doi:10.1088/1126-6708/2002/12/076
  [hep-th/0205144].
\bibitem{Bourdier:2015wda}
  J.~Bourdier, N.~Drukker and J.~Felix,
  ``The exact Schur index of $\mathcal{N}=4$ SYM,''
  JHEP {\bf 1511}, 210 (2015)
  doi:10.1007/JHEP11(2015)210
  [arXiv:1507.08659 [hep-th]].
\bibitem{Bourdier:2015sga}
  J.~Bourdier, N.~Drukker and J.~Felix,
  ``The $\mathcal{N}=2$ Schur index from free fermions,''
  JHEP {\bf 1601}, 167 (2016)
  doi:10.1007/JHEP01(2016)167
  [arXiv:1510.07041 [hep-th]].
\bibitem{Intriligator:2003jj}
  K.~A.~Intriligator and B.~Wecht,
  ``The Exact superconformal R symmetry maximizes a,''
  Nucl.\ Phys.\ B {\bf 667}, 183 (2003)
  doi:10.1016/S0550-3213(03)00459-0
  [hep-th/0304128].
\bibitem{Martelli:2005tp}
  D.~Martelli, J.~Sparks and S.~T.~Yau,
  ``The Geometric dual of a-maximisation for Toric Sasaki-Einstein manifolds,''
  Commun.\ Math.\ Phys.\  {\bf 268}, 39 (2006)
  doi:10.1007/s00220-006-0087-0
  [hep-th/0503183].
\bibitem{Franco:2005rj}
  S.~Franco, A.~Hanany, K.~D.~Kennaway, D.~Vegh and B.~Wecht,
  ``Brane dimers and quiver gauge theories,''
  JHEP {\bf 0601}, 096 (2006)
  doi:10.1088/1126-6708/2006/01/096
  [hep-th/0504110].
\bibitem{Butti:2005vn}
  A.~Butti and A.~Zaffaroni,
  ``R-charges from toric diagrams and the equivalence of a-maximization and Z-minimization,''
  JHEP {\bf 0511}, 019 (2005)
  doi:10.1088/1126-6708/2005/11/019
  [hep-th/0506232].
\bibitem{Butti:2005ps}
  A.~Butti and A.~Zaffaroni,
  ``From toric geometry to quiver gauge theory: The Equivalence of a-maximization and Z-minimization,''
  Fortsch.\ Phys.\  {\bf 54}, 309 (2006)
  doi:10.1002/prop.200510276
  [hep-th/0512240].
\bibitem{Nakayama:2005mf}
  Y.~Nakayama,
  ``Index for orbifold quiver gauge theories,''
  Phys.\ Lett.\ B {\bf 636}, 132 (2006)
  doi:10.1016/j.physletb.2006.03.045
  [hep-th/0512280].
\bibitem{Eager:2012hx}
  R.~Eager, J.~Schmude and Y.~Tachikawa,
  ``Superconformal Indices, Sasaki-Einstein Manifolds, and Cyclic Homologies,''
  Adv.\ Theor.\ Math.\ Phys.\  {\bf 18}, no. 1, 129 (2014)
  doi:10.4310/ATMP.2014.v18.n1.a3
  [arXiv:1207.0573 [hep-th]].
\bibitem{Agarwal:2013pba}
  P.~Agarwal, A.~Amariti and A.~Mariotti,
  ``A Zig-Zag Index,''
  arXiv:1304.6733 [hep-th].
\bibitem{Closset:2013vra}
  C.~Closset, T.~T.~Dumitrescu, G.~Festuccia and Z.~Komargodski,
  ``The Geometry of Supersymmetric Partition Functions,''
  JHEP {\bf 1401}, 124 (2014)
  doi:10.1007/JHEP01(2014)124
  [arXiv:1309.5876 [hep-th]].
\bibitem{Klebanov:1998hh}
  I.~R.~Klebanov and E.~Witten,
  ``Superconformal field theory on three-branes at a Calabi-Yau singularity,''
  Nucl.\ Phys.\ B {\bf 536}, 199 (1998)
  doi:10.1016/S0550-3213(98)00654-3
  [hep-th/9807080].
\bibitem{Gubser:1998fp}
  S.~S.~Gubser and I.~R.~Klebanov,
  ``Baryons and domain walls in an N=1 superconformal gauge theory,''
  Phys.\ Rev.\ D {\bf 58}, 125025 (1998)
  doi:10.1103/PhysRevD.58.125025
  [hep-th/9808075].
\bibitem{Nakayama:2006ur}
  Y.~Nakayama,
  ``Index for supergravity on AdS(5) x T**1,1 and conifold gauge theory,''
  Nucl.\ Phys.\ B {\bf 755}, 295 (2006)
  doi:10.1016/j.nuclphysb.2006.08.012
  [hep-th/0602284].
\bibitem{Martelli:2004wu}
  D.~Martelli and J.~Sparks,
  ``Toric geometry, Sasaki-Einstein manifolds and a new infinite class of AdS/CFT duals,''
  Commun.\ Math.\ Phys.\  {\bf 262}, 51 (2006)
  doi:10.1007/s00220-005-1425-3
  [hep-th/0411238].
\bibitem{Benvenuti:2005ja}
  S.~Benvenuti and M.~Kruczenski,
  ``From Sasaki-Einstein spaces to quivers via BPS geodesics: L**p,q|r,''
  JHEP {\bf 0604}, 033 (2006)
  doi:10.1088/1126-6708/2006/04/033
  [hep-th/0505206].
\bibitem{Franco:2005sm}
  S.~Franco, A.~Hanany, D.~Martelli, J.~Sparks, D.~Vegh and B.~Wecht,
  ``Gauge theories from toric geometry and brane tilings,''
  JHEP {\bf 0601}, 128 (2006)
  doi:10.1088/1126-6708/2006/01/128
  [hep-th/0505211].
\bibitem{Butti:2005sw}
  A.~Butti, D.~Forcella and A.~Zaffaroni,
  ``The Dual superconformal theory for L**pqr manifolds,''
  JHEP {\bf 0509}, 018 (2005)
  doi:10.1088/1126-6708/2005/09/018
  [hep-th/0505220].
\bibitem{Imamura:2006ub}
  Y.~Imamura,
  ``Anomaly cancellations in brane tilings,''
  JHEP {\bf 0606}, 011 (2006)
  doi:10.1088/1126-6708/2006/06/011
  [hep-th/0605097].
\bibitem{Imamura:2006ie}
  Y.~Imamura,
  ``Global symmetries and 't Hooft anomalies in brane tilings,''
  JHEP {\bf 0612}, 041 (2006)
  doi:10.1088/1126-6708/2006/12/041
  [hep-th/0609163].

\end{thebibliography}
\end{document}